\documentclass[times]{qjrms4}
\usepackage{bbm}
\usepackage[colorlinks,bookmarksopen,bookmarksnumbered,citecolor=red,urlcolor=red]{hyperref}

\newcommand\BibTeX{{\rmfamily B\kern-.05em \textsc{i\kern-.025em b}\kern-.08em
T\kern-.1667em\lower.7ex\hbox{E}\kern-.125emX}}

\usepackage{moreverb}
\def\volumeyear{0000}
\usepackage{natbib}
\usepackage[colorinlistoftodos]{todonotes}
\usepackage{multirow,bigstrut}
\newtheorem{theorem}{Theorem}[section]

\begin{document}

\runningheads{T.~T.~T.~Chau \emph{et~al.}}{Maximum likelihood estimation for nonlinear models}

\title{An efficient particle-based method for maximum likelihood estimation in nonlinear state-space models}

\author{T.~T.~T.~Chau\affil{a}\corrauth, P.~Ailliot\affil{b,c}, V.~Monbet\affil{a}, and P.~Tandeo\affil{c}}

\address{\affilnum{a} IRMAR and INRIA, Rennes, France\\
\affilnum{b} University of Western Brittany and LMBA, Brest, France\\
\affilnum{c} IMT-Atlantique and Lab-STICC, Brest, France}

\corraddr{T.~T.~T.~Chau , IRMAR, University of Rennes 1, 35042 Rennes, France.\\
Email: thi-tuyet-trang.chau@univ-rennes1.fr}

\begin{abstract}
Data assimilation methods aim at estimating the state of a system by combining observations with a physical model. When sequential data assimilation is considered, the joint distribution of the latent state and the observations is described mathematically using a state-space model, and filtering or smoothing algorithms are used to approximate the conditional distribution of the state given the observations.
The most popular algorithms in the data assimilation community are based on the Ensemble Kalman Filter and Smoother (EnKF/EnKS) and its extensions.
In this paper we investigate an alternative approach where a Conditional Particle Filter (CPF) is combined with Backward Simulation (BS). This allows to explore efficiently the latent space and simulate quickly relevant trajectories of the state conditionally to the observations.
We also tackle the difficult problem of parameter estimation. Indeed, the models generally involve statistical parameters in the physical models and/or in the stochastic models for the errors. These parameters strongly impact the results of the data assimilation algorithm and there is a need for an efficient method to estimate them. Expectation-Maximization (EM) is the most classical algorithm in the statistical literature to estimate the parameters in models with latent variables. It consists in updating sequentially the parameters by maximizing a likelihood function where the state is approximated using a smoothing algorithm. In this paper, we propose an original Stochastic Expectation-Maximization (SEM) algorithm combined to the CPF-BS smoother to estimate the statistical parameters. We show on several toy models that this algorithm provides, with reasonable computational cost, accurate estimations of the statistical parameters and the state in highly nonlinear state-space models, where the application of EM algorithms using EnKS is limited. We also provide a Python source code of the algorithm.
\end{abstract}


\maketitle

\section{Introduction} 
Data assimilation has been applied in various fields such as oceanography, meteorology or navigation \citep{ghi91,wor10,car17} to reconstruct dynamical processes given observations. When sequential data assimilation is used, a state-space model is considered. It is defined sequentially for  $t = 1:T$ by
\begin{equation} \label{eq: SSM}
\begin{cases}
x_t =  \mathcal{M}_ \theta \left(x_{t-1},~ \eta_t  \right) \\
y_t =  \mathcal{H} _ \theta \left(x_t, ~\epsilon_t \right)
\end{cases}
\end{equation}
where $(x_t, y_t)$ belong to the state and observation spaces $(\mathcal{X},\mathcal{Y})$ and $(\eta_t,\epsilon_t)$ are independent white noise sequences with covariance matrices denoted respectively $Q$ and $R$. The functions $\mathcal{M}_\theta$ and $\mathcal{H}_\theta$ describe respectively the evolution of the state $(x_t)$ and the transformation between the state and the observations $(y_t)$.  We denote $\theta \in \Theta$ the vector of parameters. For instance, $\theta$ may contain physical parameters in the models $(\mathcal{M}_\theta, \mathcal{H}_\theta)$ and error covariances $(Q,R)$.

\begin{figure*}[h]  
\centering
\includegraphics{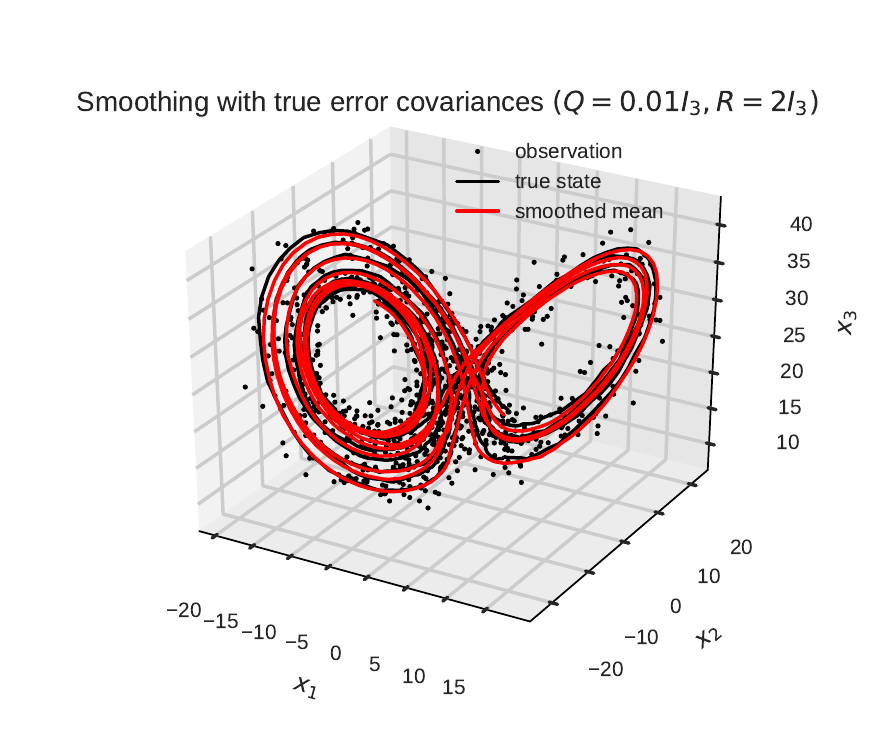}
\includegraphics{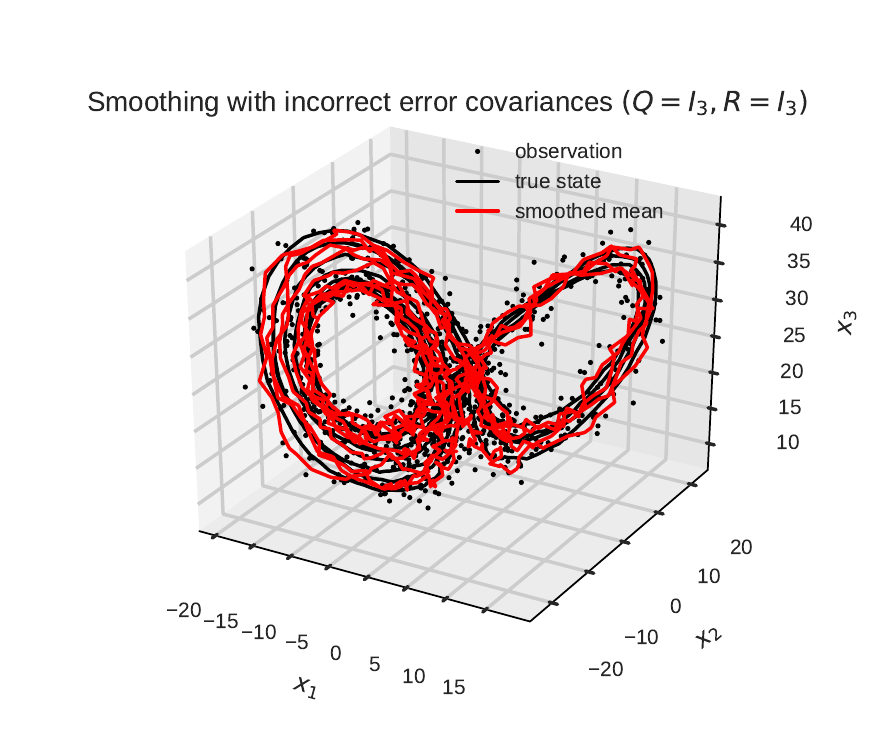}
\caption{Impact of parameter values on smoothing distributions for the Lorenz-63 model. The true state (black curve) and observations (black points) have been simulated with $\theta = (Q,R) =  (0.01I_3,2I_3)$. The mean of the smoothing distributions (read curve) are computed using a standard particle smoother \citep{dou09} with $100$ particles. Results obtained with the true parameter values $\theta^* = (0.01I_3,2I_3)$ (left panel) and wrong parameter values $\tilde{\theta} = (I_3,I_3)$ (right panel).}
\label{fig: introduction}
\end{figure*} 
Given a fixed vector $\theta$ and $T$ measurements $y_{1:T}= (y_1,...,y_T)$, data assimilation schemes relate to compute filtering distributions $\{p_ \theta (x_t|y_{1:t} )\}_{t=1:T}$ or smoothing distributions $\{p_ \theta (x_t|y_{1:T} )\}_{t=1:T}$. However, it is often difficult to identify a reasonable value of $\theta$. This is due to the diversity of observation sources, the effect of physical terms and model complexity, or numerical failures \citep{dre17,zhu17}. And incorrect values of $\theta$ may lead to bad reconstruction results. This is illustrated on Figure~\ref{fig: introduction} using the Lorenz-63 model (see Section~\ref{sec: Lor} for a formal definition). Smoothing with true parameter value provides a good approximation of the true trajectory (left panel) whereas the trajectory obtained with wrong parameter value is noisy and biased (right panel). This illustration emphasizes the role of parameter estimation in data assimilation context. A nice explanation of the problem is also given in \cite{ber13}.

One common approach to estimate parameters in data assimilation community  is based on innovation statistics \citep{ miy11,ber13, zhe15, zhu17} whose formulas were first given in \cite{des05}. Although these methods permit an adaptive estimation of the error covariances $Q$ and $R$, physical parameters of nonlinear dynamical models are difficult to estimate with this approach. An alternative is to implement likelihood-based methods. A recent review, including Bayes inference and maximum likelihood estimate, can be found in \cite{kan15}. The Bayesian approach \citep{and10, lintho13, kan15} aims to infer arbitrary parameter by simulating from the joint distribution of the state and the parameter. Additionally it is able to describe the shape of parameter distribution which might be multi-modal. In data assimilation community this approach is used in  \cite{str07,uen16,str17}. But the Bayesian approaches still have some drawbacks. First, a very large number of iterations is required to get good approximations of the parameter distributions. Moreover simulating the distributions in high-dimensional state-space models is sometimes impractical. For example, it is difficult to simulate directly the full model covariance $Q$ which involves a lot of parameters if the latent state has values in a high dimensional space. To simplify the problem, $Q$ is typically supposed to have a predefined form, such as the multiplication of a scalar by a given matrix, and only the scale factor is estimated. In this paper we hence focus on maximum likelihood estimation.

There are two major approaches in the statistical literature to maximize numerically the likelihood in models with latent variables: Gradient ascent and Expectation-Maximization (EM) algorithms. As stated in \cite{kan15} \textit{gradient ascent
algorithms can be numerically unstable as they require to scale carefully the components of the score
vector} and thence the EM approach is generally favored when considering complicated models such as the one used in data assimilation. The first EM algorithm was suggested by \cite{dem77}. Various variants of the EM algorithm were proposed in the statistical literature (see e.g. \cite{cel95,mcl07,sch11,lin13,kan15} and references therein) and in the data assimilation community (see \cite{uen14,tan15,pul17,dre17}). The common idea of these algorithms is to run an iterative procedure where an auxiliary quantity which depends on the smoothing distribution is maximized at each iteration, until some convergence criteria are reached. 

Within the EM machinery, the challenging issue is generally to compute the joint smoothing distribution $p_\theta(x_{0:T}|y_{1:T})$ of the latent state given the entire sequence of observations, where $x_{0:T} = (x_0,x_1, \cdots, x_T)$. For a linear Gaussian model, the Kalman smoother (KS) \citep{shu82} based on Rauch-Tung-Streibel (RTS) provides an exact solution to this problem. The difficulty arises when the model is nonlinear and the state does not take its values in a finite state-space. In such situation the smoothing distribution is intractable. To tackle this issue, simulation-based methods were proposed. In data assimilation, Ensemble Kalman smoother (EnKS) \citep{eve00} is one of the most favorite choices. By implementing the best linear unbiased estimate strategy, this method is able to approximate the smoothing distribution using only a few simulations of the physical model (members) at each time step. Unfortunately the approximation does not converge to the exact distribution $p_\theta(x_{0:T}|y_{1:T})$ for non-linear state-space models \citep{le09}. Particle smoothers have been proposed as an alternative in \cite{DouFreGor01, god04,CapGodMou07,dou09}. However, they demand a huge amount of particles (and thus to run the physical many times) to get a good approximation of the target probability distribution. Since 2010, conditional particle smoothers (CPSs) \citep{lin11,lin12,linsch12,lintho13,lin14,sve15}, pioneered by \cite{and10}, have been developed as other strategies to simulate the smoothing distribution. Contrary to the more usual smoothing samplers discussed above, CPSs simulate realizations using an iterative algorithm. At each iteration, one conditioning trajectory is plugged in a standard particle smoothing scheme. It helps the sampler to explore interesting parts of the state space with only few particles. After a sufficient number of iterations, the algorithm provides samples approximately distributed according to the joint smoothing distribution.

In the data assimilation community, EM algorithms have been generally used in conjunction with EnKS (EnKS-EM algorithm). Recent contributions \citep{tan15,pul17,dre17} implement this approach using $20-100$ members and concentrate on estimating the initial state distribution and error covariances. In the statistical community, the combination of standard or approximate particle smoothers (PSs) with large amount of particles and EM algorithms (PS-EM) \citep{ols08,sch11,kok14,kan15,pic18} is preferred. The number of particles is typically in the range $10^2-10^6$ which would lead to unrealistic computational time for usual data assimilation problems (the number of particles corresponds to the number of time that the physical model needs to be run at each time step). In \cite{lin13}, the author proposed to use a CPS algorithm, named Conditional particle filtering-Ancestor sampling (CPF-AS, \cite{linsch12}), within a stochastic EM algorithm (CPF-AS-SEM). The authors showed that the method can estimate $Q$ and $R$ using only $15$ particles for univariate spate-space models. However CPF-AS suffers from degeneracy (the particle set reduce to a very few effective  particles) and consequently the estimated parameters of CPF-AS have bias and/or large variance. In the present paper, we propose to combine another CPS, referred to as Conditional particle filtering-Backward Simulation (CPF-BS, \cite{lintho13}), with the stochastic EM scheme. The novel proposed maximum likelihood estimate method, abbreviated as CPF-BS-SEM, aims at estimating the parameters with only a few particles and thus reasonable computational costs for data assimilation. In this article we show that our approach has better performances than the EM algorithms combined with standard PS \citep{sch11}, CPF-AS \citep{lin13} and EnKS \citep{dre17}. Numerical illustrations are compared in terms of estimation quality and computational cost on highly nonlinear models. We also provide an open-source Python library of all mentioned algorithms which is available on-line at \url{https://github.com/tchau218/parEMDA}.

The article is organized as follows. In Section~\ref{sec:method} we introduce the main methods used in the article, including smoothing with the conditional particle smoother CPF-BS and maximum likelihood estimation using CPF-BS-SEM.  Section~\ref{sec:num} is devoted to numerical experiments and Section~\ref{sec:con} contains conclusions.

\section{Methods}
\label{sec:method}
In this section, we first introduce the conditional particle smoother which is the key ingredient of the proposed method. This  smoother is based on conditional particle filtering (CPF) which is described in Section~\ref{sec:CPF}. Standard particle filtering algorithm is also reminded and its performance is compared to the one of CPF. Section~\ref{sec:CPF filsmo} presents iterative filtering/smoothing schemes which are the combinations of CPF and ancestor tracking algorithms. We also analyze benefits and drawbacks of these filters/smoothers. Then an iterative smoothing sampler based on CPF-BS is provided as an alternative to the CPF smoothers and theirs theoretical properties are quickly discussed in Section~\ref{sec:CPFBS}. Finally, the combination of CPF-BS with the EM machinery for maximum likelihood estimation is discussed in Section~\ref{sec:EM}.

\subsection{Filtering and smoothing using conditional particle-based methods}
\subsubsection{Particle Filtering (PF) and Conditional Particle Filtering (CPF)}
\label{sec:CPF} 
In the state-space model defined by \eqref{eq: SSM}, the latent state $(x_t)_{t = 0:T}$ is a Markov process defined by its background distribution $p_\theta (x_0)$  and transition kernel $p_{\theta}(x_t|x_{t-1})$. The observations $(y_t)_{t = 1:T}$ are conditionally independent given the state process and we denote $p_{\theta} (y_t|x_t)$ as the conditional distribution of $y_t$ given $x_t$. The transition kernel $p_{\theta}(x_t|x_{t-1})$ depends on both the dynamical model $\mathcal{M}_\theta$ and the distribution of the model error $\eta_t$ whereas the conditional observation distribution  $p_{\theta} (y_t|x_t) $ is a function of the observation model $\mathcal{H}_\theta$ and the distribution of the observation error $\epsilon_t$. In this section we discuss algorithms to approximate the filtering distribution $p_{\theta}(x_{t}|y_{1:t})$ which represents the conditional distribution of the state at time $t$ given the observations up to time $t$.

For linear Gaussian models, the filtering distributions are Gaussian distributions which means and covariances can be computed using the Kalman recursions. 
When state-space models are nonlinear, as it is the typical case for data assimilation applications, the filtering distributions do not admit a closed form and  particle filtering (PF) methods have been proposed  to compute approximations of these quantities \citep{DouGodAnd98, DouFreGor01, CapGodMou07}. The general PF algorithm is based on the following relation between the filtering distributions at time $t-1$ and $t$ 
\begin{equation} \label{eq: filtering dist}
p_{\theta}(x_{0:t}|y_{1:t}) = \frac{p_{\theta}(y_t|x_{t})~ p_{\theta}(x_t|x_{t-1})}{p_{\theta}(y_{t}|y_{1:t-1})} ~p_{\theta}(x_{0:t-1}|y_{1:t-1})
\end{equation}
where $p_{\theta}(y_{t}|y_{1:t-1})$ is the normalization term of $p_{\theta}(x_{0:t}|y_{1:t})$. Note that if we are able to compute the joint filtering distribution $p_{\theta}(x_{0:t}|y_{1:t})$ then it is possible to deduce the marginal filtering distribution $p_\theta (x_{t}|y_{1:t})$ by integrating over all variables $x_{0:t-1}$. 

\begin{figure*}[h] 
\centering
\includegraphics[height = 7cm, width = 16.5cm]{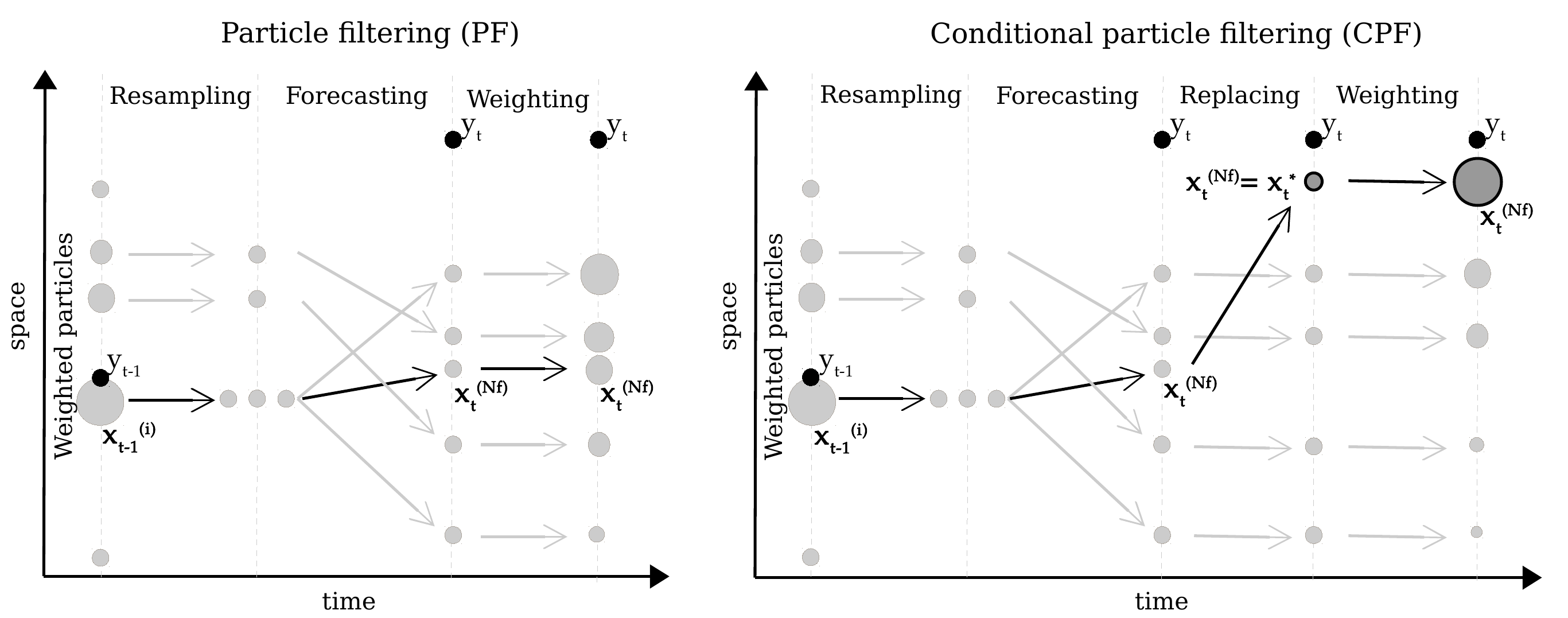} 
\centering
\caption{Comparison of PF and CPF schemes  using $N_f =5$ particles (light gray points) in time window $[t-1,t]$ on the SSM \eqref{eq: SSM}. The observation model is the identity function. The main difference is shown on black quivers as CPF replaces the particle $x_t^{(N_f)}$ with conditioning particle $x_t^*$ (dark gray point).}
\label{fig: PF vs CPF scheme}
\end{figure*}

PF runs with $N_f$ particles to approximate $p_{\theta}(x_{0:t}|y_{1:t})$ recursively in time. Let us suppose that 
initial particles $\lbrace x_{0}^{(i)} \rbrace _{i=1:N_f}$ have been drawn from the background $p_ \theta (x_0)$ and 
the filtering process has been done up to time $t-1$. Since PF is based on importance sampling, we now have a system of particles and their corresponding weights $\lbrace x_{0:t-1}^{(i)}, w_{t-1}^{(i)}\rbrace_{i = 1:N_f}$ which approximates the joint filtering distribution  $p_{\theta}(x_{0:t-1}|y_{1:t-1})$. The next step of the algorithm consists in deriving an approximation
\begin{align} \label{eq: filtering est}
\hat p_\theta \left(x_{0:t}|y_{1:t} \right) = \sum \limits _{i = 1}^{N_f} \delta_{ x_{0:t}^{(i)}} \left(x_{t} \right)w_{t}^{(i)}
\end{align}
of $p_{\theta}(x_{0:t}|y_{1:t})$ based on Eq.~\eqref{eq: filtering dist}. It 
is carried out in three main steps (see left panel of Figure~\ref{fig: PF vs CPF scheme} for an illustration): 
\begin{itemize}
\item \textbf{Resampling.} Systematic resampling method with multinomial distribution $\mathcal{M}\mathrm{ul} ( \cdot|w_{t-1}^{(1)}, w_{t-1}^{(2)}, \cdots, w_{t-1}^{(N_f)} )$ \citep{dou05,hol06} is used to reselect potential particles in $\lbrace x_{0:t-1}^{(i)} \rbrace _{i = 1:N_f}$. In this step the filter duplicates particles with large weight and removes particles with very small weight. 
\item \textbf{Forecasting.} It consists in propagating  the particles from time $t-1$ to time $t$ with a proposal kernel $\pi_{\theta} (x_t|{x_{0:t-1},y_{1:t}})$. 
\item \textbf{Weighting.} Importance weights $\lbrace w_{t}^{(i)} \rbrace _{i = 1:N_f}$ of the particles $\lbrace x_{0:t}^{(i)} \rbrace _{i = 1:N_f}$ are computed according to the formula
\begin{align} \label{eq: PF weight}
\nonumber
W \left(x_{0:t}\right) &= \frac{p_{\theta}\left(x_{0:t}|y_{1:t}\right)}{\pi _{\theta}(x_{t}|{x_{0:t-1},y_{1:t}})} \\ &\stackrel{\eqref{eq: filtering dist}}{\alpha}   \frac{p_{\theta}\left(y_t|x_{t}\right) p_{\theta}(x_{t}|x_{t-1})}{\pi _{\theta}(x_{t}|{x_{0:t-1},y_{1:t}})} ~p_{\theta}\left(x_{0:t-1}|y_{1:t-1}\right).
\end{align}
\end{itemize}
The entire algorithm of PF is presented in \textbf{Algorithm 0}. Notation $\{I_t^i\}_{t = 1:T}^{i = 1:N_f}$ in \textbf{Algorithm 0} is used to store the particle's indices accross time in order to be able to reconstruct trajectories. It is a key ingredient in the smoothing algorithms which are presented later.
Note that, in a general PF algorithm, particles can be propagated according to any proposal distribution $\pi _\theta$. If we choose $\pi_{\theta} (x_t|{x_{0:t-1},y_{1:t}}) = p_{\theta}(x_t|x_{t-1})~ p_{\theta}(x_{0:t-1}|y_{1:t-1})$ (see \cite{DouGodAnd98,pit99,CapGodMou07} and \cite{sny11} for discussions on the choice of $\pi _\theta$), the importance weight function \eqref{eq: PF weight} can be simplified as $W \left(x_{0:t}\right) ~ \alpha ~~ p_{\theta}\left(y_t|x_{t}\right)$. With this choice,  which is referred to as bootstrap filter in the literature, the forecasting step consists in sampling according to the dynamical model $\mathcal{M}_\theta$. It is the favorite choice for data assimilation applications \citep{van15,pot16,lgu17} and it is used in this article for numerical illustrations. 

Conditional particle filtering (CPF) was introduced first time by \cite{and10} and then discussed by many authors \citep{lin11,lin12,lintho13,lin14,sve15}. The main difference with PF consists in plugging a conditioning trajectory $X^* = (x_{1}^*, \cdots, x_{T}^*) \in \mathcal{X}^T$  into a regular filtering scheme. 
In practice, CPF works in  an iterative environment where the conditioning trajectory $X^*$ is updated at each iteration. This is further discussed in the next section. In this section we assume that $X^*$ is given. 
Due the conditioning, CPF algorithm differs from the PF algorithm
in adding a replacing step between the forecasting and weighting steps. In this step, one of the particles is replaced by the conditioning trajectory. It is possible to set the conditioning particle as the particle number $N_f$ and this leads to updating the position of the particles at time $t$ according to
\begin{equation} \label{eq: replacing step}
x_{t}^{(i)} =  
\begin{cases}
x_{t}^{(i)} \sim \pi_{\theta} (x_t|{x_{0:t-1}^{(I_{t}^{i})},y_{1:t}}), \quad  \forall {i=1:N_f-1}\\
x_t^*, \quad  i= N_f.
\end{cases}
\end{equation}
Similarly to PF, the reset sample $\{x_{t}^{(i)}\}_{i =1:N_f}$ is next weighted according to Eq.~\eqref{eq: PF weight}. In \textbf{Algorithm 0} we present the differences between PF and CPF algorithms. The additional ingredients of CPF are highlighted using a gray color.
  
   \textbf{\underline{Algorithm 0}: Particle Filtering (PF)/\textit{\textcolor{gray}{Conditional Particle Filtering (CPF)}} given \textit{\textcolor{gray}{the conditioning $X^* = (x_{1}^*, x_{2}^*, \cdots, x_{T}^*)$ (only for CPF)}}  and $T$ observations $y_{1:T} $, for fixed parameter $\theta$}.
 \begin{itemize}
 \item Initialization:\\
  + Sample $\{x_0^{(i)} \}_{i =1:N_f} \sim ~ p_\theta(x_0)$.\\
  + Set weights: $w _0^{(i)} =1/N_f, \forall{i =1:N_f} $.
  \item For $t =1:T$,\\
   + \textbf{Resampling}: draw indices $$\{I_t^{i}\}_{i =1:N_f} \sim \mathcal{M}\mathrm{ul} \left( j | w_{t-1}^{(1)}, w_{t-1}^{(2)}, \cdots, w_{t-1}^{(N_f)} \right)$$
   with $j \in \{1,2, \cdots,N_f\}$.\\
  + \textbf{Forecasting}: sample new particle $$x_t^{(i)} \sim \pi_{\theta}\left(x_t|x_{0:t-1}^{(I_{t}^i)}, y_{1:t} \right), \forall{i =1:N_f}.$$
  + \textit{\textcolor{gray}{ \textbf{Replacing} (only for CPF): set  $x_t^{(N_f)} = x_t^*$ and $I_t^{N_f} =N_f$.}}\\
  + \textbf{Weighting}: compute $\tilde w_t^{(i)} = W \left(x_{0:t-1}^{(I_{t}^i)},x_t^{(i)}\right)$ by using Eq. \eqref{eq: PF weight} then calculate its normalized weight $w_{t}^{(i)} = \frac{ \tilde w_{t}^{(i)}}{\sum \limits _{i = 1}^{N_f}  \tilde w_{t}^{(i)}}$, $\forall{i =1:N_f} $.
  
   end for.
  \end{itemize}
\begin{figure*}[h]
\centering
\includegraphics{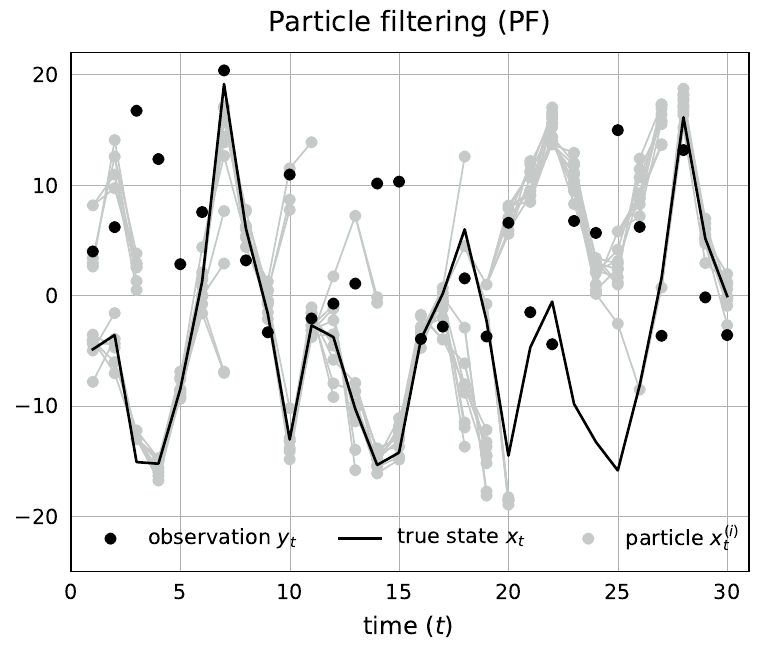} 
\includegraphics{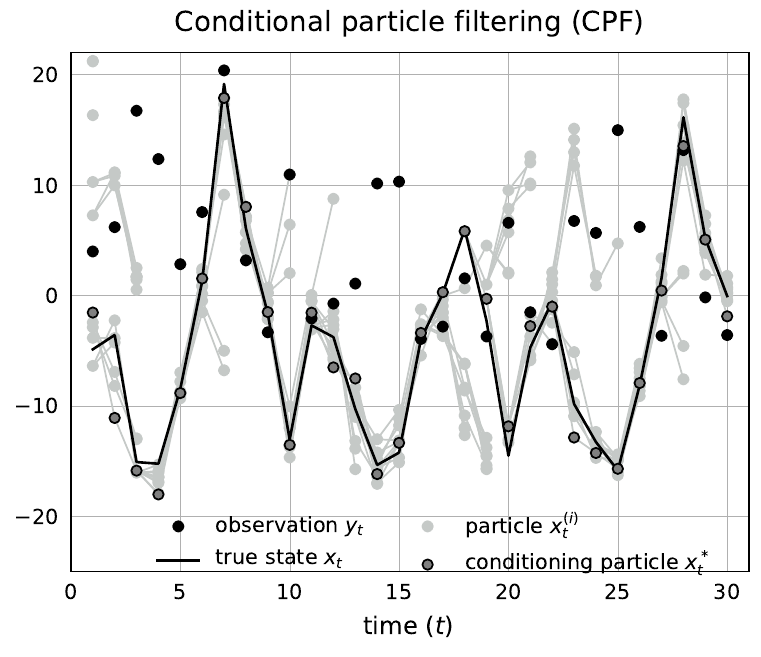} 
\centering
\caption{Comparisons of PF and CPF performances with 10 particles on Kitagawa model \eqref{eq: nonlinear SSM}, where $T =30, (Q,R) = (1,10)$. Conditioning particles (dark gray points) are supposed to live around to the true state trajectory (black curve). Gray lines are the links among particles which have the same ancestor.}
\label{fig: PF vs CPF perform}
\end{figure*}

The general principle of the CPF algorithm is also presented on Figure \ref{fig: PF vs CPF scheme}.
CPF does a selection between particles sampled from the proposal kernel $\pi_\theta$ and the conditioning particle. We can imagine two opposite situations. If the conditioning particle is "bad" (i.e. far from the true state) then the filtering procedure will eliminate it by weighting and resampling. 
But if conditioning particle is "good" (i.e. close to the true state) then it will have a high weight and it will be duplicated and propagated at the next time step. This ensures that if an interesting sequence is used as conditioning trajectory, then the CPF algorithm will explore the state space in the neighborhood of this trajectory and thus, hopefully, an interesting part of the state space. This is also  illustrated on Figure~\ref{fig: PF vs CPF perform} which has been drawn using the Kitagawa state-space model (given later in Eq.~\ref{eq: nonlinear SSM}). 
This univariate model was chosen because it is known that it is difficult to compute accurate approximations of the filtering distribution: the forecasting distribution $p_\theta (x_t|x_{t-1})$ can be bimodal due to the $\cos$-term and the observation operator is quadratic. In addition, we use a large value of $R$ to get unreliable observations. On the left panel of the figure, around time $t= 17$, PF starts to simulate trajectories which are far away from the true states. All the particles are close to $0$ and the dynamical model provides unstable and inaccurate forecasts. At the same time, the observation $y_t$ is unreliable and can not help to correct the forecasts. It leads to a bad approximation of the filtering distribution at time $t=18$. It has consequences on the next time steps: the forecast distributions remain far from the true state and the filter gives bad results. CPF gives better results thanks to a good conditioning trajectory which helps generating relevant forecasts  (see right panel of Figure~\ref{fig: PF vs CPF perform}).

When the number of particles $N_f$ is big, the effect of the conditioning particles  becomes negligible and the PF and CPF algorithms give similar results. However running a particle filter with large amount of particles is generally computationally impossible for data assimilation problems. Algorithms which can provide a good approximation of the filtering distributions using only a few particles (typically in the range $10-100$) are needed. An alternative strategy to PF/CPF with a large number of particles, based on iterating the CPF algorithm with a low number of particles, is discussed in the next section.

\subsubsection{Filtering and smoothing with conditional particle filter}
\label{sec:CPF filsmo}
A key input to the CPF algorithm is the conditioning particles of the given trajectory $X^*$. As discussed in the previous Section, the "good" conditioning particles must be "close" to the true state in order to help the algorithm simulates interesting particles in the forecasting step with reasonable computational costs. Remark also that the distribution of the particles simulated by running one iteration of the CPF depends on the distribution of the conditioning trajectory $X^*$. The distribution of $X^*$ must be chosen in such a way that the output of the CPF is precisely the filtering/smoothing distributions that we are targeting. One solution to this problem can be found in \cite{and10} (see a summary in Theorem~\ref{Theo}): if $X^*$ is simulated according to the smoothing distribution then running the CPF algorithm with this conditioning particle will provide other sequences distributed according to the smoothing distributions. A more interesting result for the applications states that if the conditioning trajectory is "bad", then iterating the CPF algorithm after a certain number of iterations will provide "good" sequences for $X^*$ which are distributed approximatively according to the smoothing distribution. At each iteration the conditioning trajectory $X^*$ is updated using one of the trajectories simulated by the CPF algorithm at the previous iteration. The corresponding procedure is described more precisely below.

\begin{figure}[h]  
\centering
\includegraphics[height = 5cm, width = 6cm]{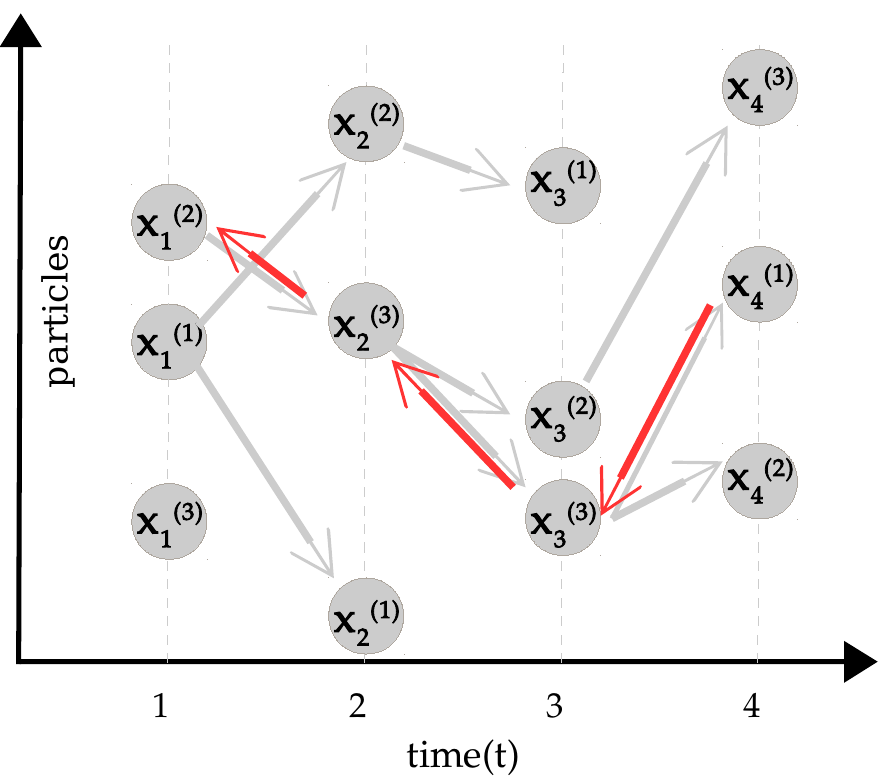} 
\centering
\caption{An example of ancestor tracking one smoothing trajectory (backward quiver) based on ancestral links of filtering particles (forward quivers). Particles (gray balls) are assumed to be obtained by a filtering algorithm with $T=4$ and $N_f = 3$.}
\label{fig: AT scheme}
\end{figure}

Running the CPF algorithm (\textbf{Algorithm 0}) until the final time step $T$ gives a set of particles, weights and indices which define an empirical distribution on $\mathcal X^{T+1}$,
\begin{equation} \label{eq: SMC}
\hat p _\theta \left(x_{0:T}|y_{1:T} \right) = \sum \limits _{i = 1}^{N_f} \delta_{ x_{0:T}^{(i)}} \left(x_{0:T} \right)w_{T}^{(i)}
\end{equation}
where $x_{0:T}^{(i)}$ is one particle path (realization) taken among particles (eg. one continuous gray link on Figure~\ref{fig: PF vs CPF perform}), $w_{T}^{(i)}$ is its corresponding weight and $i$ is an index of its particle at the final time step. The simulation of one trajectory according to Eq.~\eqref{eq: SMC}, 
is based on sampling its final particle with respect to the final weights $(w_T^{(i)})_{i =1:N_f}$ such that 
$$p(x_{0:T}^{s} = x_{0:T}^{(i)}) \propto w_T^{(i)}.$$
Then, given the final particle, eg. $x_T^s = x_T^{(i)}$, the rest of the path is obtained by tracing the ancestors (parent, grandparent, etc) of the particle $(x_T^{(i)})$. The information on the genealogy of the particles is stored in the indices $(I_t^{i})_{t =1:T}^{i = 1:N_f}$ since  $I_t^{i}$ is the index of the parent of  $x_t^{(i)}$. The technique is named ancestor tracking (also presented in statistical literature of standard PF such as \cite{doujoh09}). It is illustrated on Figure~\ref{fig: AT scheme}. Given $i = 1$, the parent of particle $x_4^{(1)}$ is the particle $x_{3}^{(I_4^{1})} = x_{3}^{(3)}$, its grandparent is the particle $x_{2}^{(I_{3}^{3})} = x_{2}^{(3)}$ and its highest ancestor is $x_{1}^{(I_{2}^{3})} = x_{1}^{(2)}$. At the end, we obtain one realization $x_{1:4}^{s} = x_{1:4}^{(1)} = (x_{1}^{(2)}, x_{2}^{(3)},x_{3}^{(3)},x_4^{(1)})$.

In practice the following procedure can be implemented to generate a path  $x_{0:T}^{s}= x_{0:T}^{(J_T)}= (x_{0}^{(J_0)}, x_{1}^{(J_1)}, \cdots, x_{T}^{(J_T)})$ according to Eq.~\eqref{eq: SMC}
	\begin{itemize}
	\item For $t =T$, draw index $J_T$ with $p(J_T = i) \propto w_T^{(i)}$ and set $x_{T}^{s} = x_{T}^{(J_T)}$.
	\item For $t< T$, set index $J_t = I_{t+1}^{J_{t+1}}$ and $x_{t}^{s} = x_{t}^{(J_t)}$.
	\end{itemize}

Finally the  iterative smoothing algorithm using CPF  can be described as follows,\\
\textbf{\underline{Algorithm 1}: Smoothing with Conditional Particle Filtering (CPF) given the conditioning $ X^* = (x_{1}^*, x_{2}^*, \cdots, x_{T}^*)$ and $T$ observations $y_{1:T}$, for fixed  parameter $\theta$}.
\begin{itemize}
\item Run CPF (\textbf{Algorithm 0}) given $X^*$ and observations $y_{1:T}$, with fixed parameter $\theta$ and $N_f$ particles.
\item Run ancestor tracking procedure $N_s$ times to simulate $N_s$ trajectories according to Eq.~\eqref{eq: SMC}.
\item Update the conditioning  particle $X^*$ with one of these trajectories.
\end{itemize} 

\begin{figure*}[h]  
\centering
\includegraphics{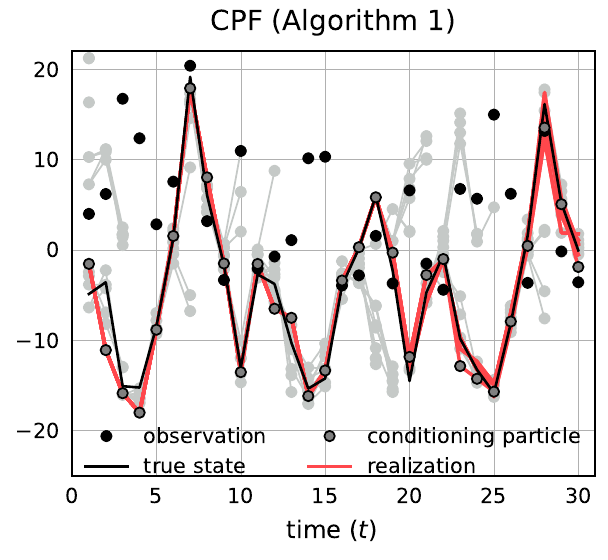} 
\includegraphics{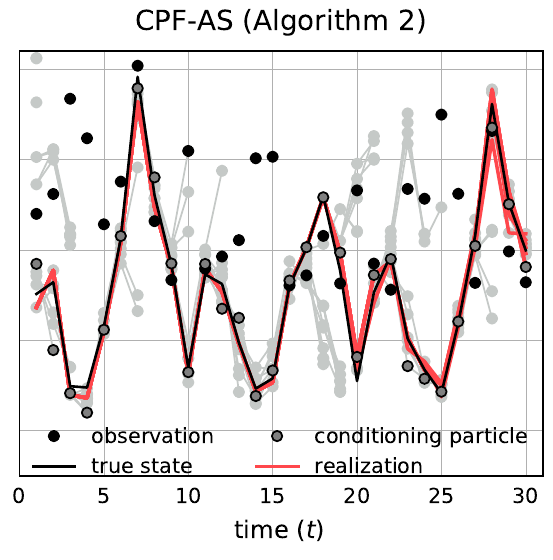} 
\includegraphics{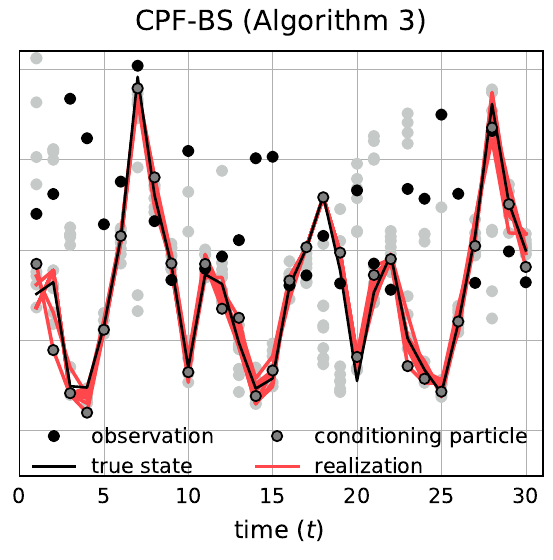} 
\centering
\caption{Comparison for simulating $N_s =10$ realizations by using CPF smoother (\textbf{Algorithm 1}), CPF-AS smoother (\textbf{Algorithm 2}) (both based on particle genealogy- light gray links) and CPF-BS smoother (\textbf{Algorithm 3}) (based on backward kernel \eqref{eq: BSS sampling est}) given the same forward filtering pattern with $N_f = 10$ particles (light gray points). The experiment is run on Kitagawa model \eqref{eq: nonlinear SSM} where $T =30, (Q,R) = (1,10)$.}
\label{fig: AS vs BSS scheme}
\end{figure*}

According to Theorem \ref{Theo} given in \cite{and10}, this algorithm will generate trajectories which are approximatively distributed according to the smoothing distribution after a certain number of iterations, even if a low number of particles is used at each iteration. However, in practice running one iteration of the CPF algorithm leads to generating trajectories which are generally almost identical to the conditioning particle  \citep{lintho13,sve15}. The main reason for this is the so-called degeneracy issue: all the particles present at the final time step $T$ share the same ancestors after a few generations. This is illustrated on Figure~\ref{fig: AT scheme}: all the  particles present at time $t=4$ have the same grandparent at time $t=2$. This is also visible on the left panel of Figure~\ref{fig: AS vs BSS scheme}. The resampling makes disappear many particles whereas other particles have many children. As a consequence, all $10$ particles at the final time step $T=30$ have the same ancestors for $t<20$. This degeneracy issue clearly favors the conditioning particle which is warranted to survive and reproduce at each time step. When iterating the CPF algorithm, the next conditioning sequence is thus very likely to be identical to the previous conditioning sequence, except maybe for the last time steps. This leads to an algorithm which has a poor mixing and lots of iterations are needed before converging to the smoothing distribution.

To improve the mixing, \cite{linsch12,lintho13,lin14} proposed to modify the replacing step of \textbf{Algorithm 0} as follows. After setting the final particle $x_t^{(N_f)} = x_t^* \in X^*$ to the conditioning particle, the index of its parent $I_{t}^{(N_f)}$ is drawn following the Bayes' rule 
\begin{align} \label{eq: AS index}
p_\theta (I_{t}^{N_f}=i|x_t^*,y_{1:t}) \propto p_\theta(x_t^*|x_{t-1}^{(i)}) ~w_{t-1}^{(i)}.
\end{align}
Resampling $I_{t}^{N_f}$ helps to break the conditioning trajectory  $X^*$ into pieces so that the algorithm is less likely to simulate trajectories which are close to $X^*$. The different steps of a smoother using this algorithm, referred to as Conditional Particle Filtering-Ancestor Sampling (CPF-AS) algorithm, are given below.\\
\textbf{\underline{Algorithm 2}: Smoothing with Conditional Particle Filtering-Ancestor Sampling (CPF-AS) given the conditioning $ X^* = (x_{1}^*, x_{2}^*, \cdots, x_{T}^* )$ and $T$ observations $y_{1:T}$, for fixed  parameter $\theta$}.
\begin{itemize}
\item Run CPF-AS, \textbf{Algorithm 0} wherein indices of conditional particles $(I_{t}^{N_f})_{t=1:T}$ are resampled with the rule \eqref{eq: AS index}, given $X^*$ and observations $y_{1:T}$, with fixed parameter $\theta$ and $N_f$ particles.
\item Run ancestor tracking procedure $N_s$ times to get $N_s$ trajectories among particles of the CPF-AS algorithm.
\item Update the conditioning  particle $X^*$ with one of these trajectories.
\end{itemize} 
In the above-mentioned references, it is shown empirically that this algorithm is efficient to simulate trajectories of the smoothing distribution with only $5-20$ particles. It is also proven that it has the same good theoretical properties (see Theorem \ref{Theo}) as the original CPF algorithm and that running enough iterations of the CPF-AS algorithm, starting from any conditioning particle $X^*$, permits to generate trajectories which are approximatively distributed according to the smoothing distribution. 

The comparison of the left and middle panels of Figure~\ref{fig: AS vs BSS scheme} shows that resampling the indices permits to obtain ancestor tracks which are different from the conditioning particles.
However, like CPF smoother (\textbf{Algorithm 1}), tracking ancestral paths in the CPF-AS smoother (\textbf{Algorithm 2}) still surfers from the degeneracy problem mentioned above. It implies that the $N_s$ trajectories simulated at one iteration of the CPF-AS generally coincide, except for the last time steps, and thus give a poor description of the smoothing distribution. This is illustrated on Figure~\ref{fig: AS vs BSS scheme}: all the trajectories simulated with the CPF-AS coincide for $t<20$ and thus can not describe the spread of the smoothing distribution.
In practice, many particles which are simulated with the physical model in the forecasting step are forgotten when running the ancestor tracking and it leads to waste information and computing resources for data-assimilation applications. In the next section we propose to replace ancestor tracking by backward simulation in order to better use the information contained in the particles.
\subsubsection{Smoothing with Conditional particle filtering- Backward simulation (CPF-BS)}
\label{sec:CPFBS}
Backward simulation (BS) was first proposed in the statistical literature in association with the regular particle filter \citep{god04, dou09, doujoh09}. Recently BS was combined with conditional smoothers \cite{lin11,lin12,lintho13}. In the framework of these smoothers, the smoothing distribution $p_\theta (x_{0:T}|y_{1:T})$ is decomposed as
\begin{equation} \label{eq: BSS dist}
p_\theta \left(x_{0:T}|y_{1:T} \right)  =  p_ \theta \left(x_{T}|y_{1:T} \right) \prod \limits _{t= 0}^{T-1} p_ \theta \left(x_{t}|x_{t+1}, y_{1:t} \right),
\end{equation}
where 
\begin{equation} \label{eq: backker}
p_ \theta(x_{t}|x_{t+1}, y_{1:t}) \propto p_ \theta \left(x_{t+1}|x_t \right)~ p_ \theta\left(x_t|y_{1:t}\right)
\end{equation}
is the so-called backward kernel.
Given the particles $ ( x_{t}^{(i)} ) _{t= 0:T} ^{i = 1:N_f} $
and the weights $ ( w_{t}^{(i)})_{t= 0:T} ^{i = 1:N_f}$ of the  CPF algorithm (\textbf{Algorithm 0}) we obtain an estimate \eqref{eq: filtering est} of the filtering distribution $p_ \theta \left(x_{t}|y_{1:t} \right)$. By plugging this estimate in \eqref{eq: backker}, we deduce the following estimate of the backward kernel
 \begin{equation}
 \label{eq: BSS sampling est}
 \hat p_ \theta \left(x_{t}|x_{t+1}, y_{1:t}\right) \propto  \sum \limits _{i = 1}^{N_f} p_ \theta(x_{t+1}| x_t^{(i)}) w_{t}^{(i)}  \delta _{x_t^{(i)}} \left( x_t \right)
 \end{equation}

\begin{figure*}[h]  
\centering
\includegraphics{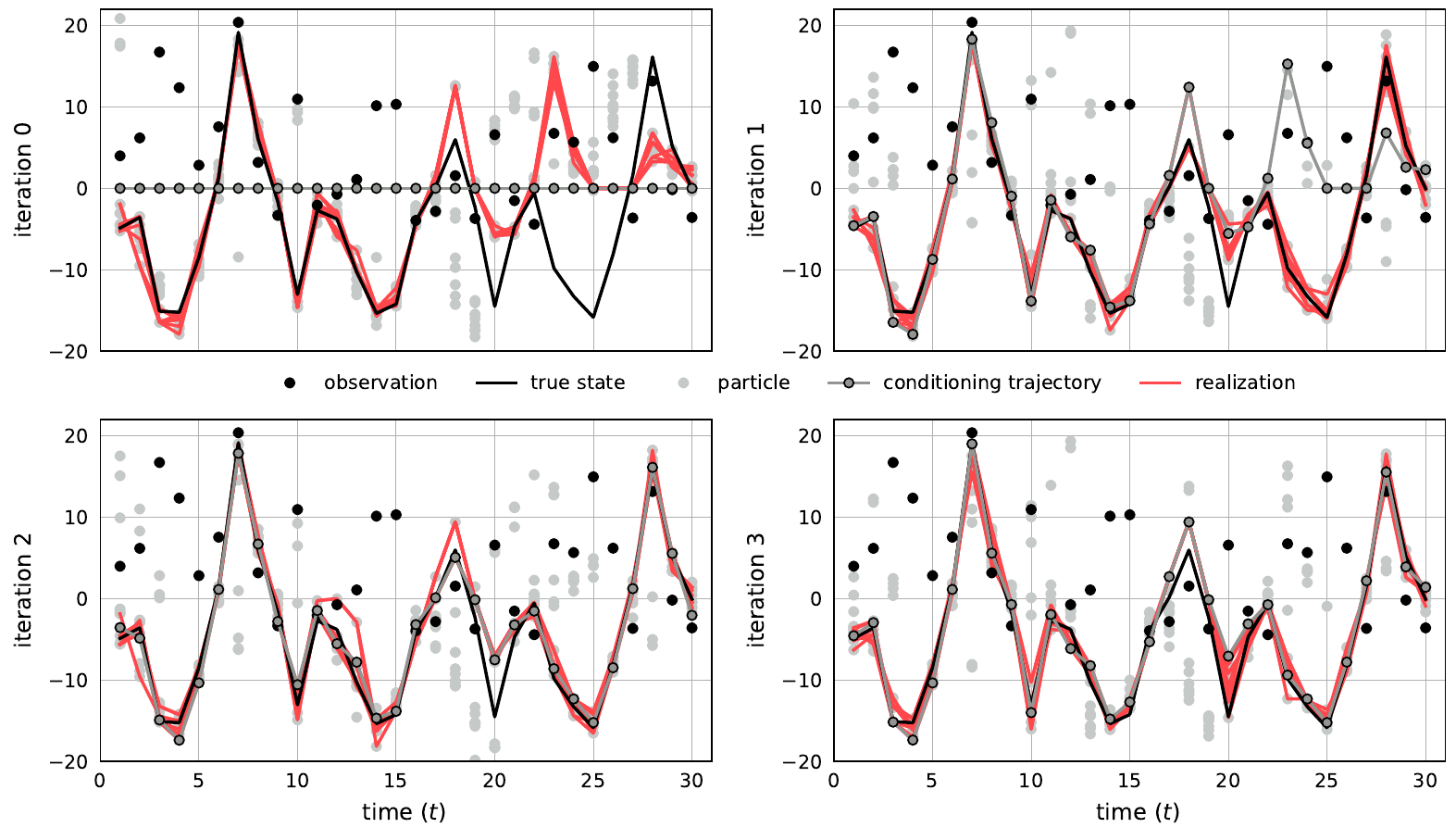} 
\centering
\caption{Performance of an iterative CPF-BS smoother (\textbf{Algorithm 3}) with $N_f = 10$ particles in simulating $N_s = 10$ realizations. The experiment is on Kitagawa model \eqref{eq: nonlinear SSM} where $(Q,R) = (1,10), T=30$. The smoother given a zero-initial conditioning ($X^* = \mathbf{0} \in \mathbb{R}^T$) is run within  $3$ iterations. For each iteration the conditioning trajectory $X^*$ is one of realizations obtained from the previous.}
\label{fig: BSS perform}
\end{figure*}

Using the relation \eqref{eq: BSS dist} and the estimate \eqref{eq: BSS sampling est}, one smoothing trajectory $x_{0:T}^{s}= x_{0:T}^{J_{0:T}} = (x_0^{(J_0)},x_1^{(J_1)},\cdots, x_{T-1}^{(J_{T-1})},x_T^{(J_T)} )$ can be simulated recursively backward in time as follows.
\begin{itemize}
  \item For $t = T$, draw $J_T$ with $p(J_T = i) \propto w_T^{(i)}$.
   \item For $t <T$,\\
  + Compute weights $w_t^{s,(i)} =  p_ \theta(x_{t+1}^{(J_{t+1})}| x_{t}^{(i)})~ w_{t}^{(i)}$ using \eqref{eq: BSS sampling est}, for all $i =1:N_f$.\\
  + Sample $J_{t}$ with $p (J_{t} = i) \propto w_{t}^{s,(i)}$.\\
  end for
  \end{itemize} 
To draw $N_s$ distinct realizations we just need to repeat $N_s$ times the procedure. The performance of BS  given outputs of one run of the CPF algorithm is displayed on Figure~\ref{fig: AS vs BSS scheme} and the complete smoother using CPF-BS is described below.\\
\textbf{\underline{Algorithm 3}: Smoothing with Conditional Particle Filtering-Backward Simulation (CPF-BS) given the conditioning $ X^* = (x_{1}^*, x_{2}^*, \cdots, x_{T}^*)$ and $T$ observations $y_{1:T}$, for fixed  parameter $\theta$}.
\begin{itemize}
\item Run CPF (\textbf{Algorithm 0}) given $( X^*, Y)$ with $N_f$ particles and fixed  parameter $\theta$.
\item Run BS procedure $N_s$ times provided the forward filtering outputs to sample $N_s$ trajectories.
 \item Update the conditioning trajectory $ X^*$ with one of these trajectories.
\end{itemize}
Figure \ref{fig: BSS perform} illustrates how the iterative CPF-BS smoother works and performs on the Kitagawa model. The smoothing procedure is initialized with a "bad" conditioning trajectory ($x^*_t=0$ for $t\in\{1,...,T\}$). This impacts the quality of the simulated trajectories which are far from the true state at the first iteration. 
Similar issues usually occurs when running regular particle smoothers (such as Particle Filtering-Backward Simulation, PF-BS, see \cite{god04,dou09}) with a small number of particles. 
The conditioning trajectory is then updated and it helps to drive the particles to interesting parts of the state space. After only $3$ iterations, the  simulated trajectories stay close to the true trajectory. Note that only $10$ particles are used at each iteration. 

\textbf{Algorithm 1}, \textbf{2} and \textbf{3} generate a new conditioning trajectory at each iteration and this defines a Markov kernel on $\mathcal X^T$ since the conditioning trajectory obtained at one iteration only depends on the conditioning particle at the previous iteration. Theorem~\ref{Theo} shows that these Markov kernels have interesting theoretical properties (see also \cite{cho15} for more results). This theorem was first proven for the CPF smoother in \cite{and10}. These results were then extended to CPF-BS in \cite{lin11} and to CPF-AS in \cite{lin14} with some extensions to solving inverse problems in non-Markovian models.
\begin{theorem} \label{Theo} 
For any number of particles $(N_f\geq 2)$ and a fixed parameter $\theta \in \Theta $,
\begin{enumerate}
\item[i.] Markov kernel $\mathcal{K}_{\theta}$ defined by one of conditional smoothers (CPF: \textbf{Algorithm 1}, CPF-AS: \textbf{Algorithm 2} and CPF-BS: \textbf{Algorithm 3}) leaves the invariant smoothing distribution $p_{\theta} (x_{0:T}|y_{1:T})$. That is, for all $X^* \in \mathcal{X}^T$ and $A \subset \mathcal{X}^{T+1}$,
 \begin{equation}
 \nonumber
 p_{\theta} (A|y_{1:T}) = \int \mathcal{K}_{\theta} (X^*,A)~ p_{\theta} (X^*|y_{1:T})~dX^*
 \end{equation}
 where $\mathcal{K}_{\theta} (X^*,A) = \mathbb{E}_{\theta, X^*}\left[\mathbbm{1}_A (x_{0:T}^{J_{0:T}}) \right]$, and $x_{0:T}^{J_{0:T}} = \{x_0^{(J_0)}, \cdots, x_T^{(J_T)} \}$.
 \item[ii.] The kernel $\mathcal{K}_\theta$ has $p_\theta$- irreducible and aperiodic. It hence converges to $p_{\theta} (x_{1:T}|y_{1:T})$ for any starting point $X^*$. Consequently,
 $$\| \mathcal{K}_\theta^{r}(X^*, \cdot) -  p_\theta (\cdot|y_{1:T})\|_{TV} \stackrel{ r \rightarrow \infty ~ \mathrm{as}}{\longrightarrow} 0.$$
 where $\| \cdot\|_{TV}$ is the total variation norm.
\end{enumerate}
\end{theorem}
The second property of this theorem implies that running the algorithm with any initial conditioning trajectory will permit to simulate samples distributed approximatively according to the smoothing distribution after a sufficient number of iterations. However, in practice, the choice of a good initial trajectory is very important, in particular when the considered state space is complex (high non-linearities, partly observed components,...). If we set an initial conditioning  trajectory far from the truth, then lots of iterations are needed before exploring a space relevant to the true state. In such situations it may be useful to provide an estimate of the true state using an alternative method (e.g. running another smoothing algorithm such as  EnKS). 

Despite of sharing the same theoretical properties as the CPF and CPF-AS smoothers, we will show in Section~\ref{sec:num} that CPF-BS algorithm gives better results in practice. This is due to its ability to avoid the degeneracy problem and hence provide better descriptions of the smoothing distribution. At the first glance, the computational cost of the backward  technique seems to be higher than the one of ancestor tracking. 
Nevertheless, for data assimilation applications, the computational complexity mainly comes from the numerical model which is used to propagate the $N_f$ particles in the forecasting step. In addition the transition probability in the backward kernel \eqref{eq: BSS sampling est} can be computed by reusing the forecast information and do not require extra runs of the physical model. The computational cost of the CPF-BS algorithm is thus similar to the ones of CPF or CPF-AS algorithms and grows linearly with $N_f$.

Recently the CPF-BS with few particles ($5-20$) has been employed to sample $\theta$ and simulate the latent state in a Bayesian framework \citep{lin11,lin12,lintho13,lin14,sve15}. 
In the next section, we propose to use the CPF-BS smoother to perform maximum likelihood estimation which is the main contribution of this paper.

\subsection{Maximum likelihood estimate (MLE) using CPF-BS}
\label{sec:EM}
In this section we discuss the estimation of the unknown parameter $\theta$ given a sequence of measurements $y_{1:T}$ of the SSM \eqref{eq: SSM}. The inference will be based on maximizing the incomplete likelihood of the observations,
\begin{equation} \label{eq: llh in smoother}
L(\theta)=p_\theta (y_{1:T}) = \int p_{\theta} \left( x_{0:T},y_{1:T} \right)  dx_{0:T}.
\end{equation}
The EM algorithm is the most classical numerical method to maximize the likelihood function in models with latent variables \citep{dem77,cel95}. It works following the auxiliary function
\begin{align} \label{eq: EM close}
G(\theta, \theta') &= \mathbb{E}_{\theta'} \left[\ln~ p_{\theta} \left(x_{0:T},y_{1:T}\right) \right] \\
&\triangleq \int \ln p_{\theta} \left(x_{0:T},y_{1:T}\right)~ p_{\theta'} \left(x_{0:T}|y_{1:T} \right) d x_{0:T}
\end{align}
where $\theta$ and $\theta'$ denote two possible values for the parameters. 
Due to Markovian assumption of the SSM \eqref{eq: SSM} and independence properties of noises $(\epsilon_t, \eta_t)$ and the initial state $x_0$, the complete likelihood $p_{\theta}(x_{0:T}, y_{1:T})$ which appears in \eqref{eq: EM close} can be decomposed as
\begin{equation} \label{eq: joint llh}
 p_{\theta}(x_{0:T}, y_{1:T}) = p_{\theta} \left( x_0 \right) ~ \prod \limits_{t =1}^T p_{\theta} \left(x_{t}|x_{t-1} \right) ~ \prod \limits_{t =1}^T p_{\theta} \left( y_{t}|x_t \right).
\end{equation}

The auxiliary function $G(.|\theta')$ is typically much more simpler to optimize than the incomplete likelihood function and the EM algorithm consists in maximizing iteratively this function. Starting from an initial parameter $\theta _0$ an iteration $r$ of the EM algorithm has two main steps:
\begin{itemize}
\item  \textbf{E-step}: compute the auxiliary quantity $G (\theta,\theta_{r-1})$
 \item \textbf{M-step}: compute $\theta_r = \arg\max \limits _{\theta}  G(\theta, \theta_{r-1})$.
\end{itemize}
It can be shown that it leads to increasing the likelihood function at each iteration and gives a sequence which converges a local maximum of $L$.


The EM algorithm combined with Kalman smoothing (KS-EM, \cite{shu82}) has been the dominant approach to estimate parameters in linear Gaussian models. In nonlinear and/or non-Gaussian models, the expectation \eqref{eq: EM close} under the distribution $p_{\theta'}(x_{0:T}|y_{1:T})$ is generally intractable
and the EM algorithm can not work in such situation. An alternative, originally proposed in \cite{cel95} and \cite{cha95}, is to use a Monte Carlo approximation of \eqref{eq: EM close}
\begin{equation} \label{eq: MCEM}
\hat G(\theta, \theta')  \triangleq \frac{1}{N_s} \sum \limits _{j = 1}^{N_s} \ln p_{\theta} \left(x_{0:T}^{j},y_{1:T}\right),
\end{equation}
where $(x_{0:T}^{j})_{j=1,...,N_s}$ are $N_s$ trajectories simulated according to the smoothing distribution $p_{\theta'} \left(x_{0:T}|y_{1:T} \right)$. This algorithm is generally named Stochastic EM (SEM) algorithm in the literature. 

To implement such procedure it is necessary to generate samples of the smoothing distribution. In the literature \citep{ols08,sch11, kok14,kan15,pic18}, standard or approximate particle smoothing methods are generally used. As discussed, it is generally computationally intractable for data assimilation applications. A classical alternative in data assimilation consists in using the EnKS algorithm \citep{eve00} leading to the EnKS-EM algorithm \citep{tan15,dre17}. Note that this procedure does not necessarily lead to increasing the likelihood function at each iteration and may not converge. Here we explore alternative procedures based on the smoothers introduced in the previous section.



\cite{lin13} proposed to use the CPF-AS smoother, leading to the CPF-AS-SEM algorithm. Given an initial parameter $\hat \theta _0$ and the first conditioning $X_0^*$, the algorithm is summed up as follows
\begin{itemize}
\item  \textbf{E-step}:
\begin{enumerate}
\item[i.] Draw $N_s$ realizations by using the CPF-AS smoother (\textbf{Algorithm 2}) once with fixed parameter $\hat \theta_{r-1}$, the conditioning $X_{r-1}^*$ and the given observations $y_{1:T}$, wherein $X^*_r$ is new conditioning trajectory obtained after updating.
\item[ii.] Compute the quantity $\hat G(\theta, \hat \theta_{r-1})$ via  Eq.~\eqref{eq: joint llh} and Eq.~\eqref{eq: MCEM}.
\end{enumerate}
\item \textbf{M-step}: Compute $\hat \theta_r = \arg\max \limits _{\theta}  ~ \hat G(\theta, \hat \theta_{r-1})$,
\end{itemize}
For each iteration $r$, $N_s$ smoothing trajectories are sampled given the previous conditioning trajectory $X_{r-1}^*$. It creates some (stochastic) dependence between the successive steps of the algorithms. This leads to such algorithm slightly different from regular EM algorithms.
In \cite{lin13} the author applied a similar algorithm to univariate models. Numerical results showed that this approach can give reasonable estimates with only few particles. Unfortunately, the degeneracy issue in the CPF-AS sampler may lead to estimates with some bias and large variance.

As discussed in the previous section, the CPF-BS smoother (\textbf{Algorithm 3}) outperforms the CPF-AS in producing better descriptions of the smoothing distribution. We hence propose a new method, CPF-BS-SEM, as an alternative to the CPF-AS-SEM for parameter estimation. The complete algorithm of the CPF-BS-SEM is presented as \\
\textbf{\underline{Algorithm 4}: Stochastic EM algorithm  using Conditional Particle Filtering-Backward Simulation (CPF-BS-SEM) given $T$ observations $y_{1:T}$.}
  \begin{itemize}
  \item Initial setting: $\hat \theta _{0}$, $X_0^*$.
  \item For iteration $r \geq 1$,\\
  + \textbf{E-step}:
  \begin{enumerate}
  \item[i.] Simulate $N_s$ samples by running CPF-BS smoother (\textbf{Algorithm 3}) once with fixed parameter $\hat \theta_{r-1}$, the conditioning $X_{r-1}^*$ and the given observations $y_{1:T}$, wherein $X^*_r$ is new conditioning trajectory obtained after updating.
  \item[ii.] Compute the quantity $\hat G(\theta, \hat \theta_{r-1})$ via  Eq.~\eqref{eq: joint llh} and Eq.~\eqref{eq: MCEM}.
  \end{enumerate}
  + \textbf{M-step}: compute $\hat \theta _r = \arg\max \limits_{\theta} \hat G(\theta, \hat \theta_{r-1})$.\\

  end for.
  \end{itemize}
 The \textbf{E-step} of this algorithm permits to get several samples at the same computational cost that the one of CPF-AS-SEM which suffers from degeneracy. That is expected to give better estimates of the quantity $G$ in Eq. \eqref{eq: MCEM}. 
Depending on the complexity of the SSM \eqref{eq: SSM}, analytical or numerical procedure may be applied in the \textbf{M-step} to maximize $\hat G$. For Gaussian state-space models the explicit expressions of estimators can be obtained directly as in the following example. Such models have popularly been considered in data assimilation context and are thus used to validate the algorithms in this article.\\
\textbf{Example}: Estimate parameter $\theta = \{Q,R \}$ in a Gaussian model 
\begin{equation} \label{eq: Gaussian SSM}
\begin{cases}
x_t =  m(x_{t-1}) + \eta_t, \quad \eta_t \sim \mathcal{N} \left( 0,Q\right)\\
y_t = h(x_t) + \epsilon _t , \quad \epsilon _t \sim \mathcal{N} \left( 0,R\right).
\end{cases}
\end{equation}
where $m$ and $h$ can be linear or nonlinear functions.\\
Through Eq.~\eqref{eq: joint llh} and \eqref{eq: MCEM}, an estimate of the function $G$ of this Gaussian model is express by
\begin{align} \label{eq: LLH function}
\nonumber
 & \hat G(\theta, \hat \theta_{r-1}) = - \frac{T}{2} \ln |Q|- \frac{T}{2} \ln |R| +  C
 \\
\nonumber
 &   - \frac{1}{2N_s} \sum \limits_{t=1}^T  \sum \limits _{j = 1}^{N_s} \left[x_t^{s,(j)}-m \left(x_{t-1}^{j} \right) \right]' Q^{-1} \left[x_t^{j}-m \left(x_{t-1}^{j} \right) \right]   \\
&   - \frac{1}{2N_s} \sum \limits_{t=1}^T  \sum \limits _{j = 1}^{N_s}  \left[y_t-h\left(x_t^{j}\right) \right]' R^{-1} \left[y_t-h\left(x_t^{j}\right) \right]
\end{align} 
where $C$ is independent to $\theta$ and $(x_t^{j} )_{t=0:T}^{j = 1:N_s}$ are sampled from the CPF-BS smoother with respect to $\hat \theta_{r-1}$. \\
Hence, an analytical expression of the estimator $\hat \theta _r = \{\hat Q_r, \hat R_r \}$ of $\theta$ which maximizes \eqref{eq: LLH function} is
\begin{align} \label{eq: ana LLH}
\nonumber
\hat Q_r  &=  \frac{1}{TN_s} ~\sum \limits_{t=1}^T \sum \limits _{j = 1}^{N_s} ~\left[x_t^{j} -m \left(x_{t-1}^{j} \right) \right] \left[ x_t^{j} -m \left(x_{t-1}^{j} \right) \right]'\\
\hat R_r  &=  \frac{1}{TN_s} ~\sum \limits_{t=1}^T \sum \limits _{j = 1}^{N_s} ~\left[y_t -h \left(x_{t}^{j} \right) \right] \left[ y_t -h \left(x_{t}^{j} \right) \right] '.
\end{align}

Different strategies have been proposed in the literature for choosing the number $N_s$ of simulated trajectories in the \textbf{E-step}. If $N_s$ is large, then the law of large numbers implies that $\hat G$ is a good approximation of $G$ and the SEM algorithm is close to the EM algorithm. It is generally not possible to run the SEM algorithm with a large value of $N_s$. In such situation, it has been proposed to increase the value of $N_s$ at each iteration of the EM (Monte Carlo EM algorithm, MCEM, see \cite{cel95}) or to re-use the smoothing trajectories simulated in the previous iterations (stochastic approximation EM algorithm, SAEM, see \cite{del99,kuh04}). It permits to decrease the variance of the estimates obtained with the SEM algorithms. For data assimilation applications, it is generally computationally infeasible to increase significantly the value of $N_s$ but the SAEM strategy could be explored. In this article, we only consider the combination of SEM and CPF-BS to facilitate the reading.

\section{Numerical illustrations} \label{sec:num}
Now we aim at validating the CPF-BS-SEM algorithm and comparing it with other EM algorithms including CPF-AS-SEM, PF-BS-SEM and EnKS-EM (such algorithms are presented in the mentioned references: \cite{ lin13, sch11} and \cite{dre17} respectively). This is done through numerical experiments on three state-space models. An univariate linear Gaussian model \eqref{eq: linear SSM} is first considered. For this model the KS-EM algorithm can be run to provide an exact numerical approximation to the MLE and check the accuracy of the estimates derived from the SEM algorithms. Next more complicated nonlinear models (Kitagawa \eqref{eq: nonlinear SSM} and a three-dimensional Lorenz-63 \eqref{eq: L63 SSM}) are considered. We focus on showing the comparisons in terms of parameter and state estimation of the CPF-BS-SEM and CPF-AS-SEM algorithms with few particles on these highly nonlinear models, where we also point out the inefficiency of the EnKS-EM algorithm.
\subsection{Linear model}
\begin{figure}[h] 
\centering
\includegraphics{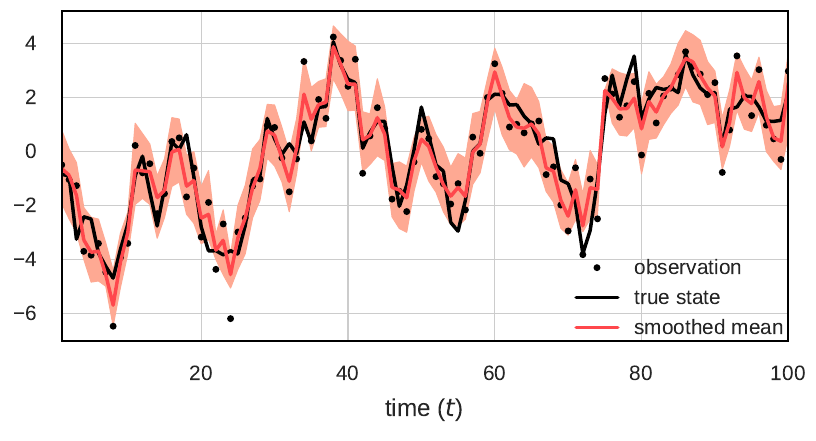} 
\centering
\caption{Reconstruction of the true state for the linear Gaussian SSM model~\eqref{eq: linear SSM} given $T=100$ observations using the CPF-BS-SEM algorithm with $10$ particles/realizations. Smoothed mean  and $95\%$ confidence interval are computed from realizations, which are simulated from last 10 iterations of the algorithm.}
\label{fig: linear-tra}
\end{figure}

A linear Gaussian SSM is defined as 
\begin{align} \label{eq: linear SSM}
\begin{cases}
x_t =  A x_{t-1} + \eta_t, \quad \eta_t \sim \mathcal{N} \left( 0,Q\right)\\
y_t = x_t + \epsilon _t , \quad \epsilon _t \sim \mathcal{N} \left( 0,R\right),
\end{cases}
\end{align}
where $(x_t,y_t)_{t =1:T} \in \mathbb{R} \times \mathbb{R}$. $\theta = (A,Q,R)$ denotes the vector of unknown parameters where $A$ is the autoregressive coefficient and $(Q,R)$ are the error variances. Implementations of stochastic version of the EM algorithms for this model are discussed in \cite{ols08,lin13,kan15}. A sequence of measurements $y_{1:T}$ is obtained by running \eqref{eq: linear SSM} with true parameter value $\theta^*= (0.9, 1, 1)$ and $T=100$ (shown on Figure~\ref{fig: linear-tra}). 
We set up the initial conditioning trajectory $X_0^*$ (only for the CPF-BS-SEM and CPF-AS-SEM algorithms) as the constant sequence equal to $0$ (the same choice is done for the models considered in Sections~\ref{sec: Kit} and \ref{sec: Lor}) and the initial parameter $\hat \theta _0$ is sampled from a uniform distribution $\mathcal{U}([0.5,~ 1.5]^3)$. 
\begin{figure*}[h] 
\includegraphics{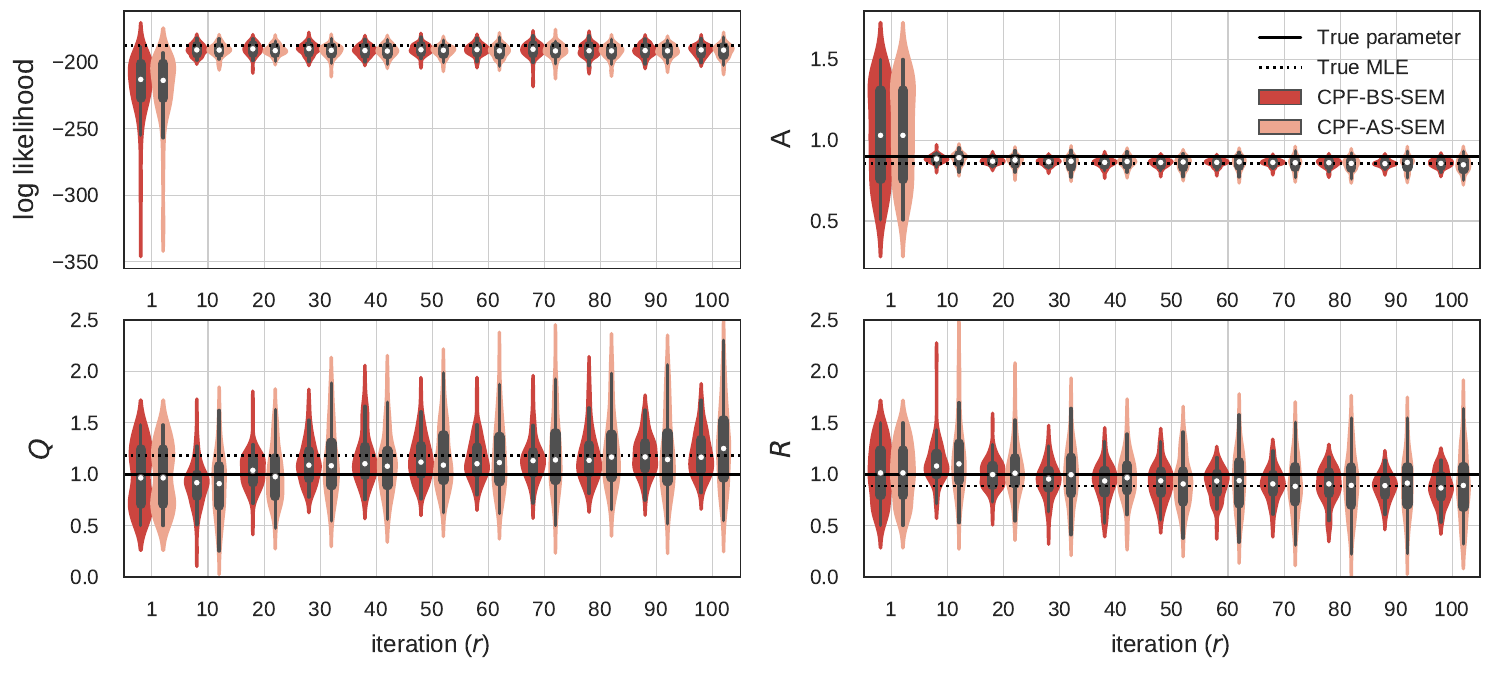} 
\centering 
\caption{Comparison between CPF-BS-SEM and CPF-AS-SEM in estimating $\theta = (A,Q,R)$ and log likelihood function for the linear Gaussian SSM model~\eqref{eq: linear SSM} with true parameter $\theta^* =  (0.9,1,1)$ and $T=100$. The results are obtained by running $100$ repetitions of the two methods with $10$ particles/realizations and $100$ iterations. The empirical distribution of the log-likelihood and parameter estimates is represented every $10$ iterations using one violin object with (black) quantile  box and (white) median point inside. The true MLE (dotted line) is computed using KS-EM with $10^4$ iterations.}
\label{fig: linear AS-Bsi}
\end{figure*}

For the first experiment, the CPF-BS-SEM and CPF-AS-SEM algorithms are run with $N_f = N_s = 10$ particles/realizations. Since the considered algorithms are stochastic, each of them is run $100$ times to show the estimators distributions. Note that in the \textbf{M-step}, the coefficient $A$ can be easily computed using Eq.~\eqref{eq: LLH function} before computing estimates of $(Q,R)$ with Eq.~\eqref{eq: ana LLH}. Figure~\ref{fig: linear AS-Bsi} shows the distribution of the corresponding estimator of $\theta$ and log likelihood every 10 iterations.  Because the model is linear and Gaussian, we can also run the KS-EM \citep{shu82} algorithm a get an accurate approximation of the true MLE of $\theta$. The estimate given by the KS-EM algorithm is shown on Figure~\ref{fig: linear AS-Bsi}. The differences with the true parameters values are mainly due to the sampling error of the MLE which is relatively important here because of the small sample size (only 100 observations to estimate 3 parameters). 
In the experiment the CPF-BS-SEM and CPF-AS-SEM algorithms start to stabilize after only 10 iterations. Even with few particles, both algorithms provide estimates which have  mean values close to the true MLE. As expected, CPF-BS-SEM is clearly better than CPF-AS-SEM in terms of variance. 
\begin{figure}[h]  
\centering
\includegraphics{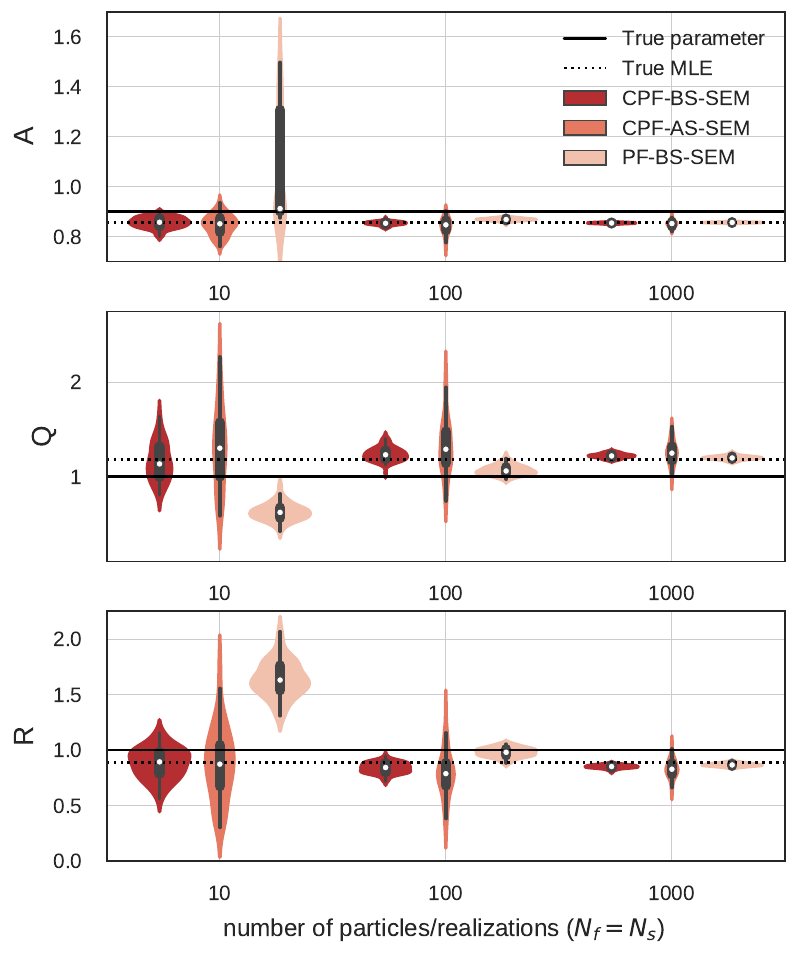} 
\caption{Comparison of the estimates of $\theta = (A,Q,R)$ at iteration $100$ of CPF-BS-SEM, CPF-AS-SEM, and PF-BS-EM  for the linear Gaussian SSM model~\eqref{eq: linear SSM} with true parameter $\theta^* =  (0.9,1,1)$ and $T=100$. These algorithms are run with different number of particles/trajectories ($N_f=N_s \in \{10,100,1000\}$). The true MLE (dotted line) is computed using KS-EM with $10^4$ iterations.}
\label{fig: linearNf}
\end{figure} 

Then we compare the CPF-BS-SEM, CPF-AS-SEM and PF-BS-SEM algorithms varying the number of  particles/realizations, $N_f = N_s \in \{10, 100, 1000 \}$. 
The empirical distribution of the final estimators $\hat \theta_{100}$ obtained by the various algorithms are shown on Figure~\ref{fig: linearNf}. The PF-BS-SEM algorithm with $N_f=N_s=10$ or even $N_f=N_s=100$ particles/realizations leads to estimates with a significant bias which is much bigger than the ones of other algorithms. It illustrates that the PF-BS-SEM algorithm based on the usual particle filter needs much more particles than the two other algorithms which use the idea of conditional particle filter. With $N_f=1000$ particles, the PF-BS-SEM and CPF-BS-SEM give similar results since the effect of conditioning becomes negligible. Then comparing the performances of the CPF-BS-SEM and CPF-AS-SEM algorithms shows again that CPF-BS-SEM is better in terms of variance. The experiment was done on different $T$-sequences of measurements and similar results were obtained. 

The reconstruction ability of the CPF-BS-SEM algorithm is displayed on Figure~\ref{fig: linear-tra}. $100$ iterations of the algorithm is run once and the $N_s = 10$ trajectories simulated in each \textbf{E-step} of the last $10$ iterations are stored. This produces $100$ trajectories. Then empirical mean and $95\%$ confidence interval (CI) of these $100$-samples are computed and plotted on Figure~\ref{fig: linear-tra}. The root of mean square error (RMSE) between the smoothed mean and the true state is $0.6996$ and the empirical coverage probability (percentage of the true states falling in the $95\%$ CIs denoted CP hereafter) is $86\%$. In theory the value should be close to $95\%$, here, the CPF-BS-SEM algorithm with non-large samples run on the short-fixed sequence of observations may give a smaller estimate of the score. An experiment to get the expected percentage is presented later (Table~\ref{Tab}).   

\subsection{Kitagawa model}
\label{sec: Kit}
\begin{figure}[h]
\centering
\includegraphics{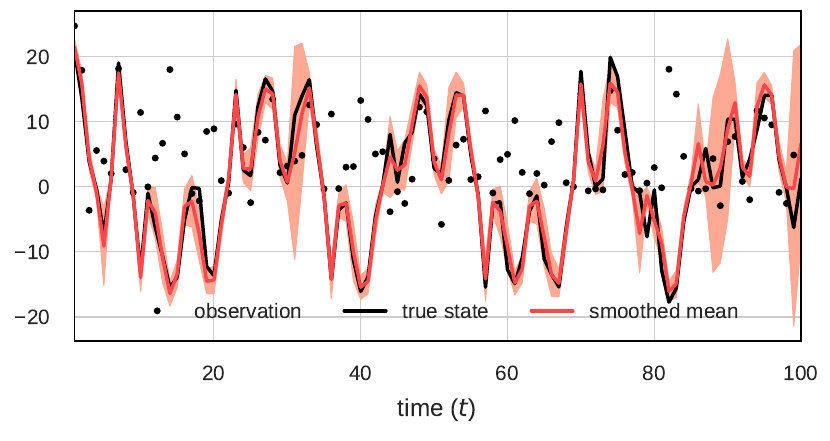} 
\centering
\caption{Reconstruction of the true state using CPF-BS-SEM with $10$ particles/realizations on the Kitagawa model \eqref{eq: nonlinear SSM} given $T=100$ observations. Smoothed means and $95\%$ confidence intervals of all realizations simulated from last 10 iterations of the algorithm are presented.}
\label{fig: kita-tra}
\end{figure}

The algorithms are now applied on a highly nonlinear system widely considered in the literature to perform numerical illustrations on SSM \citep{kit98, DouGodAnd98,god04,sch11,kok14}. Both $m$ and $h$ of the model are nonlinear and defined as follows
\begin{equation} \label{eq: nonlinear SSM}
\begin{cases}
x_t =   0.5x_{t-1} + 25 \frac{x_{t-1}}{1+x_{t-1}^2}+ 8 \cos(1.2t) + ~ \eta_t, ~~ \eta_t \sim \mathcal{N} \left( 0,Q\right)\\
y_t = 0.05x_t^2 + ~  \epsilon _t,~~ \epsilon _t \sim \mathcal{N} \left( 0,R\right)
\end{cases}
\end{equation}
where $(x_t,y_t)_{t=1:T} \in \mathbb{R} \times \mathbb{R}$. We denote  $\theta = (Q,R)$ the unknown parameter.
One sequence of $T =100$ observations generated with true parameter value $\theta^* = (1, 10)$ is shown on Figure~\ref{fig: kita-tra}. Similar values are used in \cite{god04}. The large value of the observation variance $R$ leads to generate low quality observations and thus complicate the inference.  Using only these 100 observations $y_{1:100}$, the target is to estimate $\theta$ and the true state $x_{1:100}$. The initial parameter value is simulated according to the uniform distribution $\hat \theta_0 \sim \mathcal{U} ([1, 10]^2)$. 
\begin{figure}[h]
\centering
\includegraphics{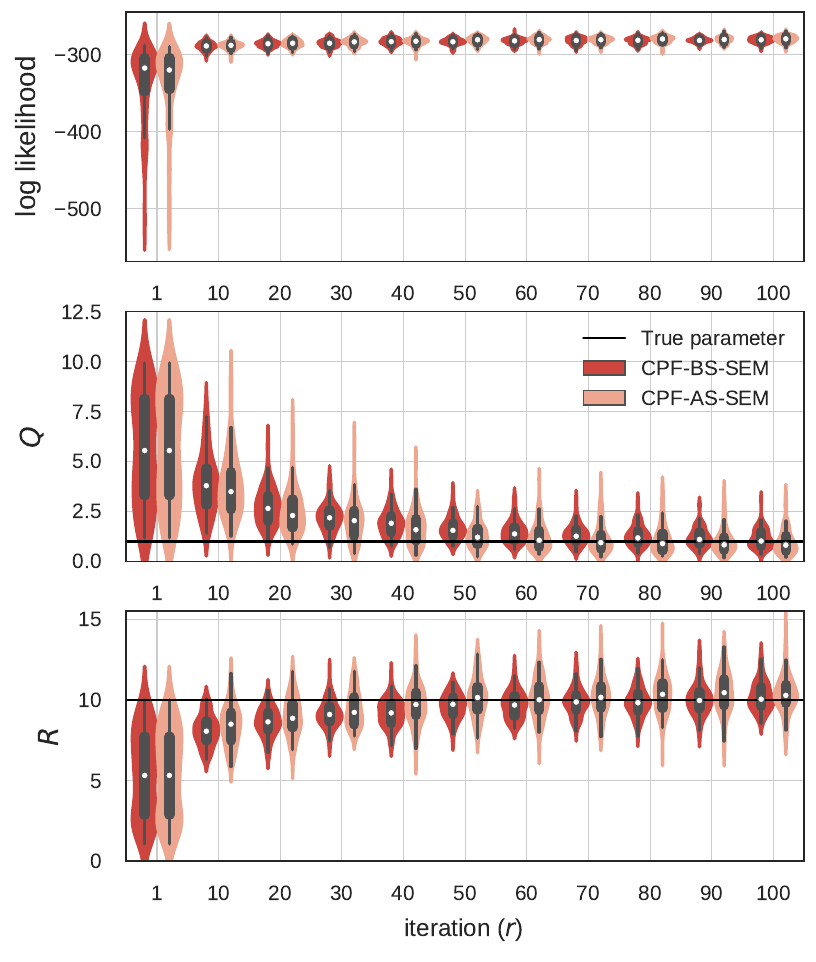}
\centering
\caption{Comparison of the CPF-BS-SEM and CPF-AS-SEM algorithms on the Kitagawa model \eqref{eq: nonlinear SSM}, where true parameter is $\theta^* = (1,10)$ and number of observations is $T=100$. The results are obtained by running $100$ times of these methods with $10$ particles/realizations and $100$ iterations. The empirical distribution of the log-likelihood and parameter estimates is represented every $10$ iterations using one violin object with (black) quantile box and (white) median  point inside.}
\label{fig: kita As-Bsi}
\end{figure} 

In this section, we only compare the CPF-BS-SEM and CPF-AS-SEM algorithms since PF-BS-SEM can not work with a small number of particles (as shown in the linear case) and \cite{lin13} also illustrated that CPF-AS-SEM using $N_f=15$ particles outperforms PF-BS-SEM using $N_f= 1500$ particles and $N_s = 300$ realizations on the Kitagawa model. In the first experiment CPF-BS-SEM and CPF-AS-SEM are run with $N_f = N_s = 10$ particles/realizations. A comparison of the two methods in terms of estimates of log likelihood and parameter $\theta = (Q,R)$ is shown in Figure \ref{fig: kita As-Bsi}. Even with few particles the estimates obtained with the two methods seem to stabilize after $50$ iterations and again 
the CPF-BS-SEM algorithm permits to reduce  the variance of the estimates compared to  the CPF-AS-SEM algorithm .
\begin{figure}[h]
\centering
\includegraphics{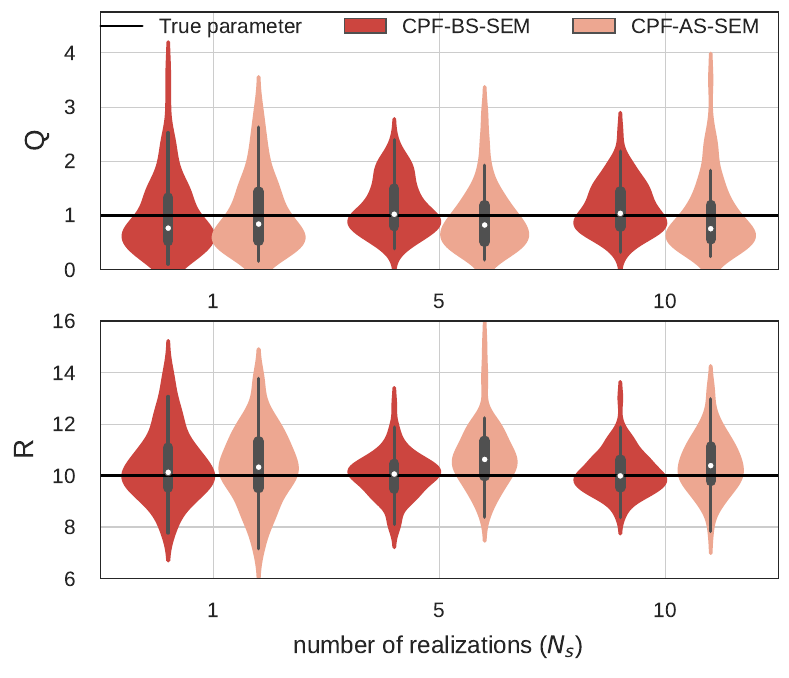}
\centering
\caption{Comparison of the estimates of $\theta = (Q,R)$ at iteration $100$ of the CPF-BS-SEM and CPF-AS-SEM algorithm on the Kitagawa model \eqref{eq: nonlinear SSM}, where true parameter is $\theta^*= (1,10)$ and number of observations is $T=100$. The algorithms are run with fixed number of particles ($N_f =10$) and different number of trajectories ($N_s \in \{1,5,10 \}$).}
\label{fig: kitaNs}
\end{figure} 

In the second experiment we run the two algorithms with fixed number of particles ($N_f = 10$) but different numbers of realizations $(N_s \in \{1, 5, 10\})$. Figure \ref{fig: kitaNs} displays the corresponding empirical distributions of $\hat \theta_{100}$. It shows that CPF-AS-SEM gives almost the same distributions of estimates as CPF-BS-SEM with $N_s = 1$. Moreover CPF-AS-SEM could not improve the estimate when we increase $N_s$ because of the degeneracy issue. CPF-BS-SEM with $N_s =5$ and $N_s =10$ gives better estimates in terms of bias and variance.  In practice it seems useless to use a large value of $N_s$ when using BS given forward filtering information. Here CPF-BS-SEM with $N_s =5$ has similar performance as CPF-BS-SEM with $N_s =10$ (see also \cite{god04,lintho13,sve15}).

Figure \ref{fig: kita-tra} shows the results obtained when reconstructing the latent space using the CPF-BS-SEM algorithm (using the same approach than for the linear model, based on storing the sequences simulated in the last 10 iterations of the algorithm). The mean of the empirical smoothing distribution seems to be close to the true state. The width of the confidence intervals varies in time and is larger (eg. at $t \in [85, 90]$) when the true state is more difficult to retrieve from the observations. The RMSE and the empirical CP with respect to the empirical smoothing distribution are $2.2478$ and $84\%$.
\subsection{Lorenz-63 model} 
\label{sec: Lor}
\begin{figure}[h]
\centering
\includegraphics{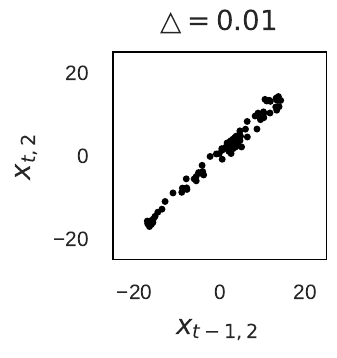}
\includegraphics{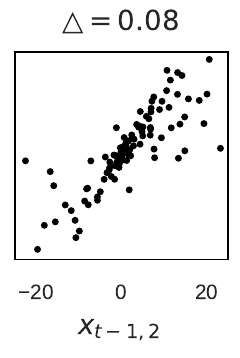}
\includegraphics{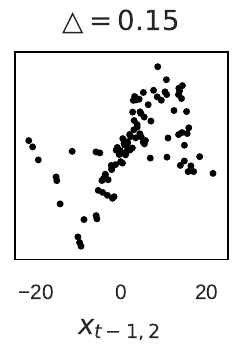}\\
\includegraphics{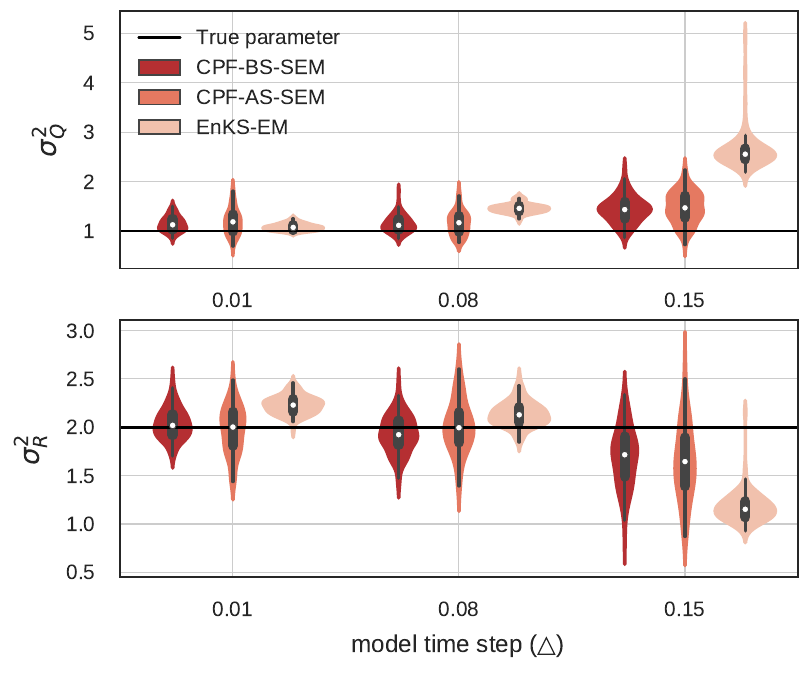}
\centering
 \caption{Comparison of the estimates of $\theta = (\sigma_Q^2, \sigma_R^2)$ for the CPF-BS-SEM, CPF-AS-SEM and EnKS-EM algorithms with $20$ members/particles for the Lorenz-63 models~\eqref{eq: L63 SSM} with varying model time step $\triangle  \in \{0.01,0.08, 0.15\}$, true parameter $\theta^* = (1,2)$ and number of observations is $T=100$. First row: scatter-plot $(x_{t-1,2},x_{t,2})$ of the unobserved variable  with respect to each model step $(\triangle)$. Last two rows: empirical distribution of the estimates of $\theta$ computed using $100$ repetitions of each algorithm at the final iteration $r =100$.}
\label{fig: L63 model step}
\end{figure}

In this section we consider the Lorenz-63 model where only the first and last components are observed. This is one of the favorite toy models in data assimilation because it is a complex (nonlinear, non-periodic, chaotic) but low-dimensional (3 components) dynamical system. More precisely, the considered state-space model is defined as 
\begin{equation} \label{eq: L63 SSM}
\begin{cases}
x_t =  m(x_{t-1}) +  \eta_t, \quad \eta_t \sim \mathcal{N} \left( 0,Q\right)\\
y_t = 
\begin{bmatrix}
    1 & 0 & 0 \\
    0 & 0 & 1 
\end{bmatrix}
  x_t +  \epsilon _t, \quad  \epsilon _t \sim \mathcal{N} \left( 0,R\right),
\end{cases}
\end{equation}
with $(x_t,y_t)_{t=1:T} \in \mathbb{R}^3 \times \mathbb{R}^2$. 
The dynamical model $m$ is related to the \cite{lor63} model  defined through the ordinary differential equation (ODE) system,
\begin{equation} \label{eq: L63 ODE }
\frac{dx_\tau}{d\tau} = g(x_\tau),
\end{equation}
where $g(x)  =\left(10(x_2-x_1), ~ x_1(28-x_3) -x_2,~  x_1x_2 - 8/3 x_3 \right)$, $~ \forall x=(x_1,x_2,x_3)' \in \mathbb{R}^3$.
In order to compute $m(x_{t-1})$, we run a Runge-Kutta scheme (order 5) to integrate the system~\eqref{eq: L63 ODE } on the time interval $[0,\triangle]$ with initial condition $x_{t-1}$. The value of $\triangle$ affects the non-linearity of the dynamical model $m$. According to Figure~\ref{fig: L63 model step} (see top panels), when  $\triangle=0.01$ the relation between $x_{t-1}$ and $x_{t}$ is well described by a linear model whereas when $\triangle=0.15$ the dynamical model has more time to evolve between successive observations and non-linearities are more pronounced. The intermediate value $\triangle=.08$ corresponding to $6$-hour recorded-observed data in atmospheric application was considered in the works of \cite{TanAil15,lgu17} and \cite{dre17}. 

For the sake of simplifying illustrations, error covariances are assumed to be diagonal. More precisely we assume that $Q = \sigma_Q^2 I_3$ and $R = \sigma_R^2 I_2$ and the unknown parameter to be estimated is $\theta = (\sigma_Q^2, \sigma_R^2) \in \mathbb R^2$. Note that an analytical solution can be derived for the \textbf{M-step} of the EM algorithm in this constrained model. 
It leads to the following expression for updating the parameters in the iteration $r$ of the EM algorithm $$\hat \theta_r = \left(\hat \sigma_{Q,r}^2, \hat \sigma_{R,r}^2 \right) = \left(\frac{\mathrm{Tr}[ \hat Q_r]}{3}, \frac{\mathrm{Tr} [\hat R_r]}{2} \right)$$ where $\hat Q_r$ and $\hat R_r$ come from \eqref{eq: ana LLH}.  The initial parameter value of the EM algorithm is drawn using a uniform distribution $\hat \theta_0 \sim \mathcal{U} ([0.5, 2] \times [1, 4])$. 
\begin{figure}[h]
\centering
\includegraphics{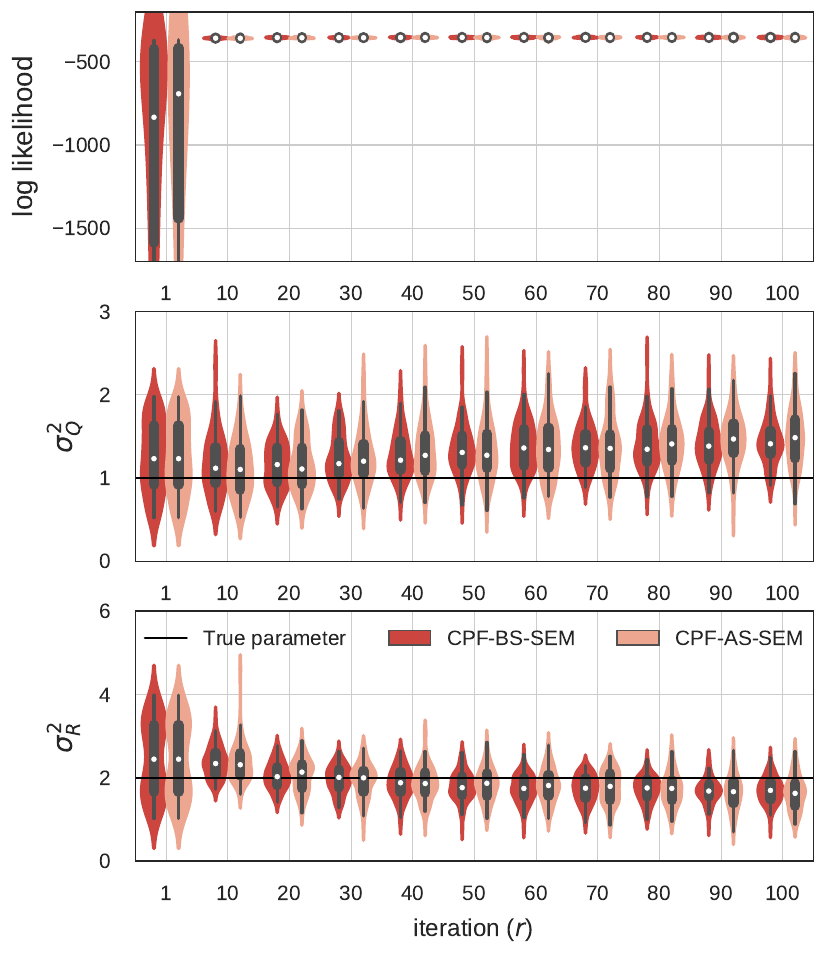}
\centering
\caption{Comparison 
between CPF-BS-SEM and CPF-AS-SEM on Lorenz-63 models~\eqref{eq: L63 SSM} with model time step $\triangle =0.15$, true parameter $\theta^* = (1,2)$ and $T=100$ observations. Results obtained by running $100$ repetitions of these methods with $20$ particles/realizations and $100$ iterations. The empirical distribution of the log-likelihood and parameter estimates is represented every $10$ iterations using one violin object with (black) quantile  box and (white) median  point inside.}
\label{fig: L63 As-Bsi}
\end{figure}

\begin{figure}[h] 
\centering
\includegraphics{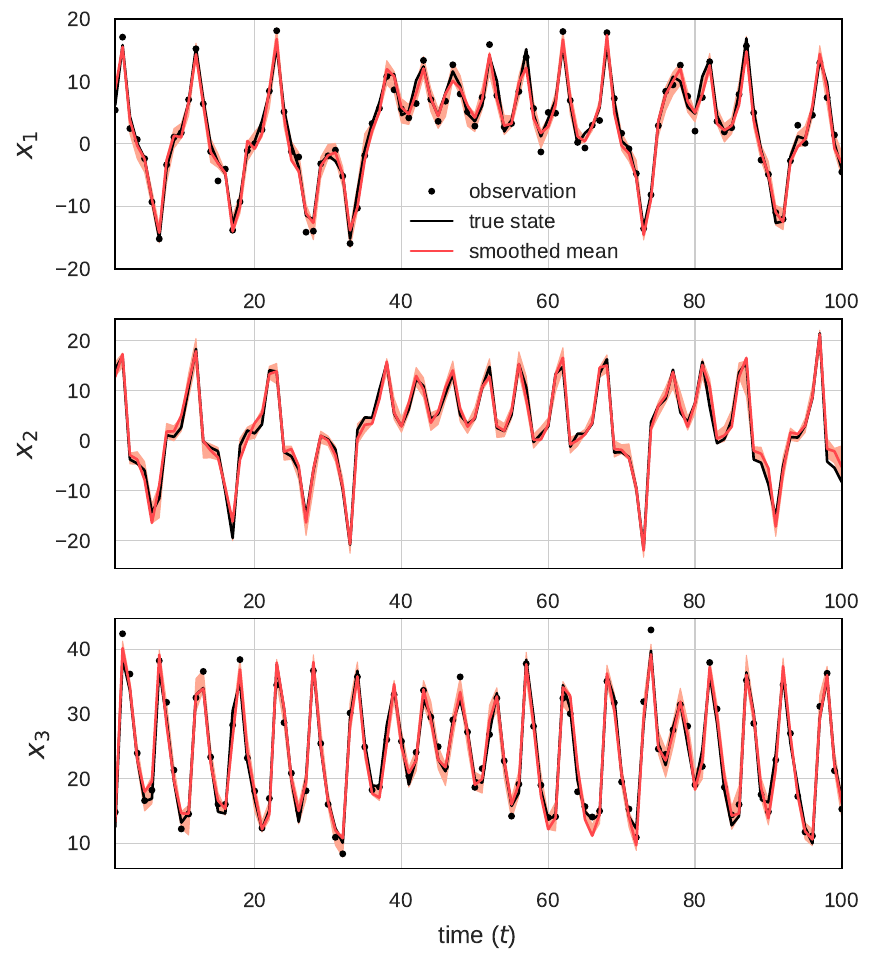} 
\caption{Reconstruction of the true state for Lorenz-63 model~\eqref{eq: L63 SSM} with $\triangle = 0.15, T =100$ by using the CPF-BS-SEM algorithm with $20$ particles/realizations. Smoothed mean and $95\%$ confidence interval of all realizations of last 10 iterations of the algorithm are computed.}
\label{fig: L63-tra}
\end{figure}

For the first experiment we simulate $T = 100$ observations of the Lorenz-63 model \eqref{eq: L63 SSM} with the model time step $\triangle = 0.15$ (it corresponds to around $20$ loops of the Lorenz-63 system) and true parameter $\theta^* = (1, 2)$ (shown on Figure~\ref{fig: L63-tra}). 
The CPF-BS-SEM and CPF-AS-SEM algorithms are compared
on Figure~\ref{fig: L63 As-Bsi}. With only $N_f=N_s=20$ particles/realizations, the two methods provide reasonable estimates of the parameters.
The comparison has been done in different scenarios, with varying true parameter values $\theta^*$, and similar results were obtained. A lower number of particles and realizations (eg. $N_f=N_s =10$) can be used in these SEM algorithms but more iterations are needed (eg. $200$) to obtain appropriate conditioning trajectories. 

In the second experiment, we also compare the results obtained with the ones of the EnKS-EM algorithm. The EnKS-EM algorithm with a low number $N$ of members often gets numerical issues when computing empirical covariances. Values of $N$ in the range $[20,1000]$ has been chosen in lots of data assimilation schemes using EnKS \citep{eve00,uen10,uen14,tan15,lgu17,dre17,pul17}. We have chosen to run the  three algorithms with $20$ members/particles to have comparable computational costs.
The experiment is run on different simulated sequences of length $T =100$, where the model time step in~\eqref{eq: L63 SSM} varies  $\triangle  \in \{0.01,0.08, 0.15\}$. 
According to Figure~\ref{fig: L63 model step}, the CPF-BS-SEM algorithm gives better estimates compared to the CPF-AS-SEM and EnKS-EM algorithms. The bias and variance of the estimates obtained with the three algorithms increase with  $\triangle$ and the non-linearity of the dynamic model. Note that the discrepancy increases quicker for the EnKS-EM algorithm. We found that it completely fail when $\triangle =0.25$ whereas the CPF-BS-SEM and CPF-AS-SEM  algorithms still give reasonable estimates  (not shown; the Python library is available for such tests). This illustrates that the EnKS-EM algorithm is less robust to non-linearities compared to the two algorithms based on the conditional particle filter.
Figure~\ref{fig: L63-tra} shows the results obtained when reconstructing the latent space using the CPF-BS-SEM algorithm (using the same approach as for the linear model, based on storing the sequences simulated in the last 10 iterations of the algorithm). The smoothed means of three variables are close to the true state and RMSEs for each component are $(0.8875, 1.0842, 1.2199)$. $95\%$ CIs covers the true state components with respect to CPs $(87\%, 84\%, 88\%)$. Although the second variable $x_{2}$ is unobserved the algorithm provides a reasonable reconstruction of this component. 

\begin{table}[h] 
\centering
\caption{Comparison of the reconstruction quality between the CPF-BS and CPF-AS smoothers on a test sequence  in terms of root of mean square error-(RMSE) and coverage probability (CP). The parameters are  estimated on a sequence of length $T=100$ (mean values of the final estimates shown on Figure~\ref{fig: L63 As-Bsi}). The CPF-BS and CPF-AS algorithms are run on a test sequence simulated using the Lorenz-63 model~\eqref{eq: L63 SSM} with $\triangle  =0.15, T'=1000, \theta^* = (1,2)$. The two scores are computed on the second component of the samples drawn from these smoothers with $20$ particles/realizations.}
\label{Tab}
\begin{tabular}{||c|c|c|c|c|c||}
\hline \hline 
\multicolumn{2}{||c|}{number of iterations} & $5$ & $10$ & $50$ & $100$ \\ \hline
\multirow{2}{*}{\bf CPF-BS}& RMSE & $1.5310$ & 1.2507 & 1.0098 & 0.9891 \\ \cline{2-6}					
	& CP &	$83.8\%$ & $88.6\%$ & $94.3\%$ & $95.7\%$ \\
 \hline
 \multirow{2}{*}{\bf CPF-AS}& RMSE & $2.1595$ & $1.5711$ & $1.0125$ & $0.9769$  \\ \cline{2-6}
	& CP  & $58.9\%$ & $78.5\%$ & $92.0\%$ & $94.8\%$ \\ 
 \hline \hline
\end{tabular}
\end{table}
Finally we preformed a cross-validation exercise to check the out-of-sample reconstruction ability of the proposed method. First we compute the mean values of the final estimates of the CPF-BS-SEM and the CPF-AS-SEM shown on Figure~\ref{fig: L63 model step}.  This provides point estimates for the parameters based on a first sequence of $T=100$ observations. Then the CPF-BS and CPF-AS algorithms  are run on an another (test) sequence of observations of length $T'=1000$ of the Lorenz-63 model with the same parameter values. This provide an estimate of the smoothing distribution for the test sequence. Table~\ref{Tab} gives RMSEs and CPs for the unobserved component of all smoothing samples with respect to number of iterations in $\{5,10,50,100\}$. As expected the CPs of the two algorithms tend to $95\%$ when the number of samples is large enough. The CPF-BS smoother clearly outperforms the CPF-AS as it gets smaller RMSEs and larger CPs with small number of iterations and thus less computational cost. Similar conclusions hold true when comparing the scores for the first and third components.

\section{Conclusion}
\label{sec:con}
CPF-BS and CPF-AS algorithms permits to simulate conditional trajectories of the latent state given observations with a low number of particles (eg. $5-20$). That allows to apply  particle filtering (smoothing) on data assimilation context. Compare to EnKS, these algorithms permits to consider highly nonlinear and/or non-Gaussian state-space models. The CPF-BS sampler leads to a better description of the smoothing distribution at the same computational cost as the CPF-AS ones which only permits to generate one trajectory. Combined with EM methodology, it provides an efficient method to estimate the parameters such as error covariances. It also permits a better estimation of the uncertainty on the reconstructed space in data assimilation. 

\end{document}